# Resignificando la tecnología para las lenguas indígenas

*U ka'a tsolik le tecnología uti'al le máasewal t'aano'obo'*

*Redefining technology for indigenous languages*

Silvia Fernández Sabido,[1] Laura Peniche Sabido[2]

[1]Centro de Investigación en Ciencias de Información Geoespacial, Yucatán, México

[2]Educativa México, Mérida, México

**RESUMEN**

En este trabajo ofrecemos un panorama de las lenguas indígenas, identificamos las causas de su desvalorización y la necesidad de legislar sobre derechos lingüísticos. Hacemos un recorrido de las tecnologías usadas para revitalizar estas lenguas, encontrando que, cuando vienen del exterior, frecuentemente tienen el efecto contrario al que buscan; pero, cuando son elaboradas desde adentro de las comunidades, se convierten en potentes instrumentos de expresión. Proponemos que la inclusión de conocimientos indígenas en los grandes modelos de lenguaje (LLM), enriquecerá el panorama tecnológico, pero debe realizarse en un entorno participativo donde se fomente el intercambio de saberes.

**Palabras clave:** lenguas indígenas, descolonización, tecnología lingüística, inteligencia artificial.

# KÓOMTS'ÍIB

Te' ju'una', k ts'áaik jump'éel tsoolil yóok'ol le máasewalo'ob t'aano'obo', k identificar ba'ax beetik u p'áatal ma' u tojol, yéetel identificar k'a'abéet u beeta'al a'almaj t'aano'ob yóok'ol u páajtalil le t'aano'obo'. K xak'altik le tecnologías ku meyaj uti'al u ka'a kuxkíinta'al le t'aano'oba', k ilik ken u taalo'ob táanxel lu'umil, tu menudo yaan u efecto opuesto ti' le ba'ax ku kaxtiko'ob; Ba'ale' ken u beeta'alo'ob ichil le kaajo'obo', ku suuto'ob nu'ukulo'ob yaan u muuk'il uti'al u ya'ala'al ba'alo'ob. K tukultike' u táakbesa'al u k'ajóolil máasewalo'ob ti' nukuch modelos t'aano'ob (LLMs) yaan u ayik'alkúunsik le paisaje tecnológico, ba'ale' k'a'ana'an u beeta'al ti' jump'éel entorno participativo tu'ux ku ts'a'abal u muuk' u ts'aiko'on k'ajóolil.

**Ba'axob yóok'olal ku t'aan:** máasewáal t'aano'ob, déeskolonisasyoon, téeknolojiya ti' liingwiistikáa, íintelijeensya áartifisyaal.

# ABSTRACT

In this paper, we offer an overview of indigenous languages, identifying the causes of their devaluation and the need for legislation on language rights. We review the technologies used to revitalize these languages, finding that when they come from outside, they often have the opposite effect to what they seek; however, when developed from within communities, they become powerful instruments of expression. We propose that the inclusion of Indigenous knowledge in large language models (LLMs) will enrich the technological landscape, but must be done in a participatory environment that encourages the exchange of knowledge.

## INTRODUCCIÓN

En este estudio llevamos a cabo una revisión bibliográfica sobre los principales esfuerzos que se han llevado a cabo para integrar las lenguas y culturas de los pueblos originarios al ámbito tecnológico. Nuestro objetivo es, con base en las experiencias pasadas, poder anticipar los riesgos y beneficios de la integración de las lenguas indígenas a la inteligencia artificial (IA), específicamente, a los grandes modelos de lenguaje, donde actualmente no están representadas. El documento está organizado como sigue: primero, presentamos una serie de reflexiones preliminares sobre lo que significa ser indígena, el valor de los conocimientos ancestrales y el papel de las lenguas; segundo, describimos la metodología usada en la revisión; tercero, mostramos una actualización del panorama global de las lenguas originarias en cuanto a hablantes, vulnerabilidad, enseñanza y marcos legales; cuarto, recorremos los diferentes estadios de la tecnología analizando los impactos reportados en la vida de las comunidades; cerramos con las conclusiones y propuestas acerca de la integración de las lenguas indígenas a la tecnología actual.

## REFLEXIONES PRELIMINARES

Iniciamos este trabajo reflexionando sobre lo que significa ser indígena -a través de algunas preguntas que el tema sugiere por sí mismo-, pues esta cuestión plantea una serie de paradojas: ¿todos hemos sido indígenas?, ¿todos somos indígenas o podemos serlo? Esta condición carece de límites claros y puede dar lugar a confusiones identitarias. Diversas definiciones[1] coinciden en que los indígenas son los descendientes de pueblos que habitaban territorios antes de ser colonizados. Desde esta perspectiva, todos somos indígenas pues en algún punto de nuestra genealogía encontraremos a un ancestro

---

[1] Batalla, "El concepto de indio en América"; Ivers, "Pueblos indígenas"; Johansson, "Miguel León-Portilla y el mundo indígena"; ONU, "Pueblos indígenas | Naciones Unidas"; OIT, "Convenio C169 - Convenio sobre pueblos indígenas y tribales".

perteneciente a un pueblo originario.

En la actualidad, la identificación de una persona como indígena suele basarse en su lugar de residencia, su lengua, vestimenta y costumbres, así como en su situación de marginación. Sin embargo, surgen preguntas interesantes: ¿un indígena deja de serlo si se traslada a una ciudad y adopta un estilo de vida diferente? ¿Qué diríamos si este individuo olvida o deja de usar su lengua originaria y adopta la cultura de un nuevo entorno? Cuando una persona indígena migra de un entorno rural a uno urbano se genera un proceso de adaptación en el que se mezclan elementos de la ruralidad con las nuevas experiencias de la ciudad, reconfigurando la identidad.[2]

Hoy día se estima que menos del 6% de la población mundial se identifica como indígena,[3] lo que sugiere que cerca del 94% de la humanidad se ha alejado de su identidad originaria en diversos grados. La condición indígena parece ser un un espectro que incluye desde individuos que mantienen una conexión profunda con la naturaleza, sus costumbres y su lengua, hasta los que han perdido su herencia cultural. De aquí surgen otras preguntas: ¿Se puede recuperar el indigenismo perdido? ¿Es posible reconectar con la naturaleza aprendiendo la lengua de nuestros ancestros y los conocimientos asociados? Tal vez una gran parte del rescate de lenguas y culturas se encuentre en ese reencuentro.

Por otro lado, podemos hablar de los conocimientos y valores asociados a las culturas indígenas. Aun cuando la vida de los pueblos originarios está permeada por tradiciones heredadas de la época colonial y moderna, podemos percibir un rico tapiz de saberes e idiomas ancestrales[4] que revela su capacidad de sintetizar grandes campos de observación creando lo que ciertos autores denominan "sistemas de conocimiento indígena", que abarcan a la naturaleza, lo económico, lo filosófico, la salud, la tecnología, la organización social, el orden económico, la filosofía de la cultura, la gobernanza y el entorno educativo.[5] Hoy en día, estos pueblos conservan la habilidad de leer señales de anticipación climática, siguen interpretando signos del cuerpo ante una variedad de enfermedades y preservan formas de producción y alimentación orgánicos. Conocimientos que son resultado de miles de años de coevolución con su entorno, y que además presentan pocas externalidades negativas y requieren pocos insumos.[6]

---

[2] Corona, "La identidad indígena como identidad urbana. Un abordaje descolonial a las crónicas de Ana Matías Rendón"; Garcés Velásquez, "Las comunidades virtuales del quichua ecuatoriano".
[3] Ivers, "Pueblos indígenas".

Aunque no todos los indígenas son conscientes de ese conocimiento, encontramos su impronta en las lenguas que siguen hablando. Por ejemplo, en el maya se tienen nombres para todas las partes del cuerpo humano, externas e internas, por pequeñas que sean, y ese saber es útil para médicos y parteras, tanto indígenas como no indígenas.[7] Asimismo, algunos pueblos utilizan formas de comunicación únicas, como silbidos o tañido de tambores que imitan el habla, con los cuales transmiten mensajes a larga distancia aprovechando las propiedades bioacústicas del entorno.[8] Ejemplos de ello son los silbos gomeros de las Islas Canarias, España, o el de la comunidad mazateca que habita el norte del estado de Oaxaca, México, quienes han desarrollado sus propios sistemas de comunicación empleando chiflidos con los que comparten mensajes complejos que viajan kilómetros resonando a través de la accidentada geografía de sus territorios. Así, las lenguas, además de ser instrumentos de comunicación para sus hablantes, son vehículos de cultura, epistemologías y modos de vida.[9]

---

[4] Mager et al., "Challenges of language technologies for the indigenous languages of the Americas"; Makgopa, "Implications of the Fourth Industrial Revolution (4IR) on the Development of Indigenous Languages of South Africa".

[5] Ajani et al., "Revitalizing Indigenous Knowledge Systems via Digital Media Technologies for Sustainability of Indigenous Languages"; Ibáñez Blancas et al., "El cambio climático y los conocimientos tradicionales, miradas desde Sudamérica"; Tshifhumulo y Makhanikhe, *Handbook of Research on Protecting and Managing Global Indigenous Knowledge Systems*.

[6] Deance Bravo y Troncoso, "TotoOffice"; Ibáñez Blancas et al., "El cambio climático y los conocimientos tradicionales, miradas desde Sudamérica"; Valladolid y Sandoval Chayña, "PERÚ".

[7] Martín Briceño y Briceño Chel, *Manual de Comunicación para Médicos. Español-maya.*; Castillo Tec, *U áanalte'il u tsikbalil ts'aak. Manual de frases médicas.*; Janetsky, "Estas parteras tradicionales combinan la herencia maya con la medicina occidental para salvar vidas".

[8] Meyer, Dentel, y Seifart, "A methodology for the study of rhythm in drummed forms of languages".

Las culturas originarias albergan saberes que merecen ser preservados y explorados, y la lengua es, quizá, la mejor forma de acceder a ellos. La interacción "lengua-sociedad" plantea todo un campo de investigación para las ciencias sociales y uno de sus principales retos es comprender todo lo que se pierde con la extinción de una lengua.[10] Conscientes de este perjuicio, diversos actores han emprendido iniciativas para frenar y revertir la disminución de hablantes y frecuentemente usan la tecnología para potenciar sus acciones.

**METODOLOGÍA**

Realizamos la búsqueda bibliográfica en motores académicos (Dimensions-AI, ERIC, Scielo, Redalyc y Google Scholar) usando las fórmulas: "tecnología"+"lenguas indígenas"; "*indigenous languages*"+"*technology*"; "*native languages*"+"*technology*"; "*natural language processing*"+"*indigenous languages*". La generalidad de estas pautas busca capturar una amplia gama de tecnologías lingüísticas y el contexto en que se produjeron. Para la última fórmula priorizamos los resultados de 2020 a la fecha, ya que se trata de un área relacionada con la IA, que presenta un avance acelerado.

Descargamos los datos en el gestor bibliográfico Zotero, resultando en 76 elementos de tipo: libro, artículo científico, artículo de conferencia, artículo periodístico, tesis, informe, estatuto/ley, blog y página web. Los criterios de exclusión fueron: la repetición de ideas y el enfoque a lenguas nativas no indígenas. Por el origen de las autoras, aunque el enfoque es global, existe un marcado énfasis en América Latina y específicamente, en México.

---

[9] Corona, "La identidad indígena como identidad urbana. Un abordaje descolonial a las crónicas de Ana Matías Rendón".
[10] Ibáñez Blancas et al., "El cambio climático y los conocimientos tradicionales, miradas desde Sudamérica"; Ríos Hernández, "El Lenguaje".

Con el fin de contextualizar nuestros hallazgos, primero analizamos el estado global de las lenguas indígenas según: el número de hablantes, los factores que las amenazan, los esfuerzos de preservación a través de la enseñanza, la brecha tecnológica y los marcos legales aplicables a la protección lingüística. Seguidamente, presentamos la comparativa de los impactos positivos y negativos de las diferentes tecnologías que se han utilizado para estudiar, difundir y/o preservar las lenguas indígenas. Finalizamos con ejemplos de activismo digital indígena y el estado de las lenguas originarias en el panorama de la IA. Con este análisis, buscamos respaldar las buenas prácticas en el diseño de tecnologías futuras.

**PANORAMA DE LAS LENGUAS INDÍGENAS**

**Número de hablantes**

El Atlas Mundial de las Lenguas[11] reporta la existencia de más de 8 mil idiomas, de los cuales, cerca de 7 mil siguen en uso. Aún cuando es difícil establecer cuántas lenguas se hablan en cada punto del planeta, pudimos darnos una idea de su distribución consultando fuentes variadas. La Figura 1.a muestra los porcentajes aproximados de lenguas habladas por continente, donde observamos que Asia, con unos 50 países donde se hablan más de 2300 lenguas, es el continente más poblado y con mayor diversidad lingüística del planeta. Están presentes desde algunas lenguas nativas de Siberia, hasta las más habladas en el mundo, como el chino o el hindi.[12] Por su parte, África está formado por 54 países y más de 3000 comunidades étnicas que hablan cerca de 2 mil lenguas, en su mayoría originarias como el suajili, el igbo y el yoruba.[13] En lo que respecta a Oceanía, es territorio de 1500 lenguas autóctonas y, tan sólo en Papúa Nueva Guinea, se estima la pervivencia de más de 800.[14] América es hogar de aproximadamente 1000 lenguas nativas:[15] 35 en Canadá, 155 en Estados Unidos[16] y el resto en Latinoamérica,[17] donde México, con 68 lenguas originarias oficiales,[18] encabeza la lista.[19] Finalmente, en Europa hay alrededor de 250 lenguas, algunas habladas por grupos minoritarios como los sami y los nenets del Ártico[20].

---
[11] UNESCO, "The World Atlas of Languages".
[12] ITACAT, "Las lenguas de Asia"; Roche et al., "The Politics of Fear and the Suppression of Indigenous Language Activism in Asia".

La Figura 1.b muestra que, paradójicamente, el 6% de la población mundial que se identifica como indígena, habla una abrumadora cantidad de 4 mil lenguas, es decir, casi el 60%.[21] Este mismo grupo pertenece al 20% más pobre, con una esperanza de vida 20 años menor que los no indígenas; al mismo tiempo, resguardan el 80% de la biodiversidad, así como los sistemas de conocimiento asociados a sus lenguas.[22]

**Estados de vulnerabilidad**

El mosaico multicultural y lingüístico que nos define como humanos, se encuentra amenazado ante el predominio de unas cuantas lenguas y culturas. El rezago tecnológico de las poblaciones originarias ha creado barreras que dificultan la transferencia del conocimiento entre los pueblos.[23] El Atlas de Lenguas de la Unesco[24] clasifica los idiomas según su grado de vulnerabilidad en: a salvo o dominante, si prevalece en los dominios públicos formales, o en riesgo, si puede caer en desuso. A esta última categoría se le asignan varios niveles que pueden verse en la Figura 1.c.

---

[13] Tierney, Ercikan, y Rizvi, "Diversity, Democracy, and Social Justice in Education: Africa".
[14] WOLACO, "Familias de lenguas en Oceanía"; Koller y Thompson, "The Representation of Indigenous Languages of Oceania in Academic Publications".
[15] Campbell, *American Indian Languages*; Mager et al., "Neural Machine Translation for the Indigenous Languages of the Americas".
[16] Peterson y Zepeda, "Towards an Indigenously-informed Model for Assessing the Vitality of Native American Languages".
[17] Ivers, "Pueblos indígenas".
[18] DOF, Programa Especial de los Pueblos Indígenas.
[19] Mager et al., "Neural Machine Translation for the Indigenous Languages of the Americas"; Mcquown, "The indigenous languages of latin america".
[20] Amiel, "¿Quiénes son los pueblos indígenas de Europa y cuáles son sus luchas?"; IIWGIA, "IWGIA Annual Report"; Council of Europe, "European Day of Languages".
[21] ONU, "Foro permanente para las cuestiones indígenas - Documento de antecedentes"; Ajani et al., "Revitalizing Indigenous Knowledge Systems via Digital Media Technologies for Sustainability of Indigenous Languages".
[22] Ivers, "Pueblos indígenas".
[23] Sandoval-Forero, "Hitos Demográficos del Siglo XXI".
[24] UNESCO, "The World Atlas of Languages".

Observamos que de las 8324 lenguas, habladas o de señas, documentadas por gobiernos, instituciones públicas y comunidades académicas, sólo 65 (0.8%) son lenguas dominantes. Del resto, 1181 (14.2%) están extintas y más de 7 mil presentan algún grado de vulnerabilidad. En este grupo se encuentran la mayoría de las lenguas indígenas. Es importante señalar que estos estados pueden cambiar con el tiempo por las acciones en favor o en contra de su preservación.

En una entrada de video, la lingüista Lindsay Williams[25] explica cómo, la clasificación de lenguas (e.g. en riesgo, minoritaria, minorizada, indígena, oficial, regional, criolla, etc.), es confusa porque se sobreponen social y lingüísticamente. Por ejemplo, el guaraní de Paraguay es una lengua indígena en riesgo pero no es minoritaria (más del 80% de la población la habla) y es una de las lenguas oficiales de Paraguay y Bolivia. El galés es otro ejemplo, pues se trata de una lengua nativa oficial en Reino Unido y, aunque minoritaria, no es considerada en peligro grave después de algunas décadas de revitalización. Otros ejemplos son el hausa de Nigeria (lengua indígena, no minoritaria, ni en peligro) y el kristanga de Malasia (lengua criolla producto de la mezcla entre portugués y malayo, minoritaria y en peligro). En nuestro contexto mexicano, el maya yucateco es una lengua indígena, no minoritaria que está en riesgo.[26]

---

[25] *What Are Minority, Indigenous, and Endangered Languages?*
[26] *What Are Minority, Indigenous, and Endangered Languages?*

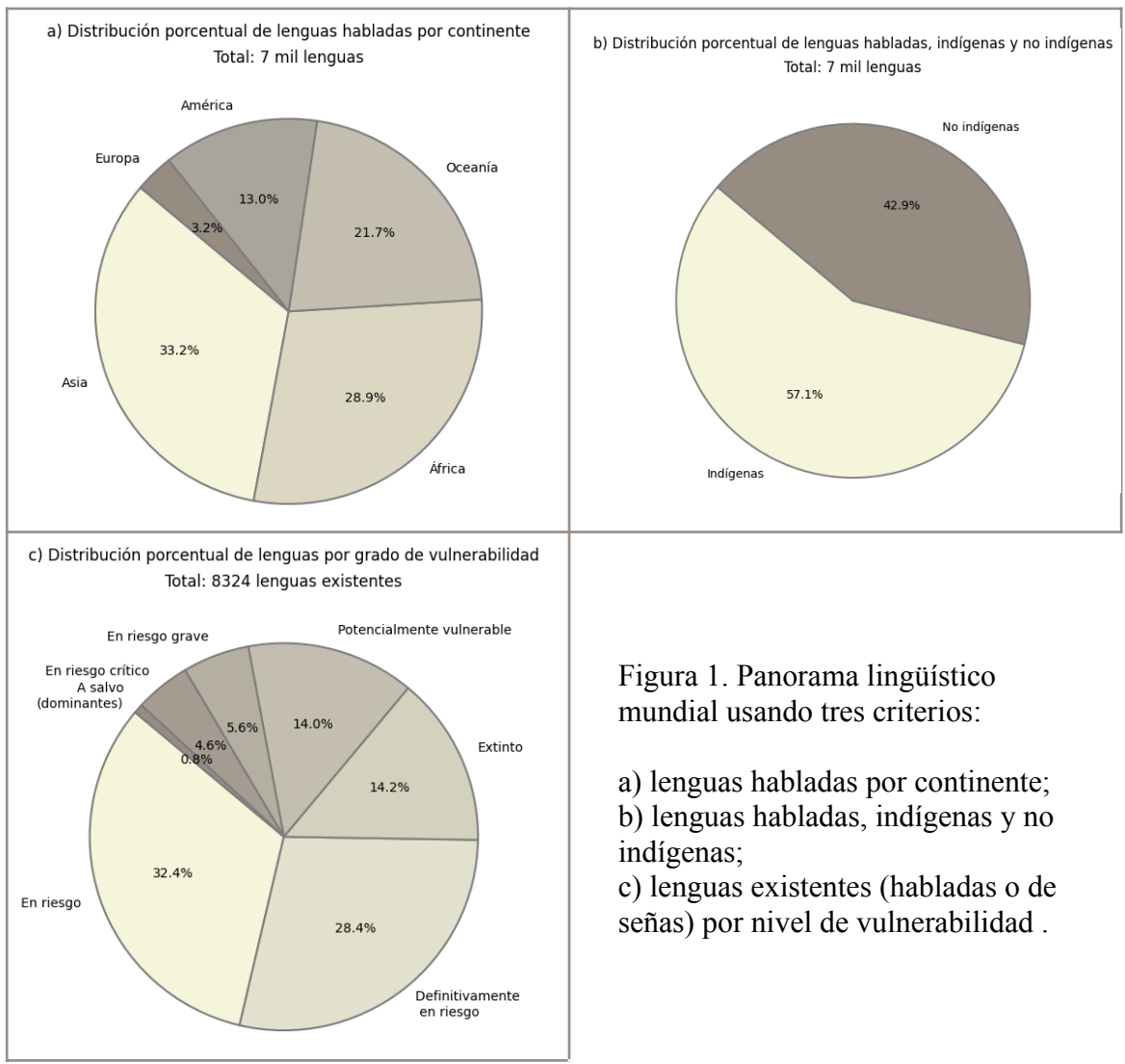

Figura 1. Panorama lingüístico mundial usando tres criterios:

a) lenguas habladas por continente;
b) lenguas habladas, indígenas y no indígenas;
c) lenguas existentes (habladas o de señas) por nivel de vulnerabilidad.

De acuerdo con Peterson y Zepeda,[27] actualmente la mitad de las lenguas indígenas del mundo ya no son habladas por niños, lo que las pone en riesgo. Algunas de las principales causas de la pérdida lingüística están esquematizadas en la Figura 2. Vemos que la monocultura es uno de las factores principales. Muchas lenguas indígenas están en peligro de extinción debido a la imposición de la cultura moderna desde la colonización, situación agravada por la globalización, que amenaza la diversidad a través de la estandarización en

---

[27] Peterson y Zepeda, "Towards an Indigenously-informed Model for Assessing the Vitality of Native American Languages".

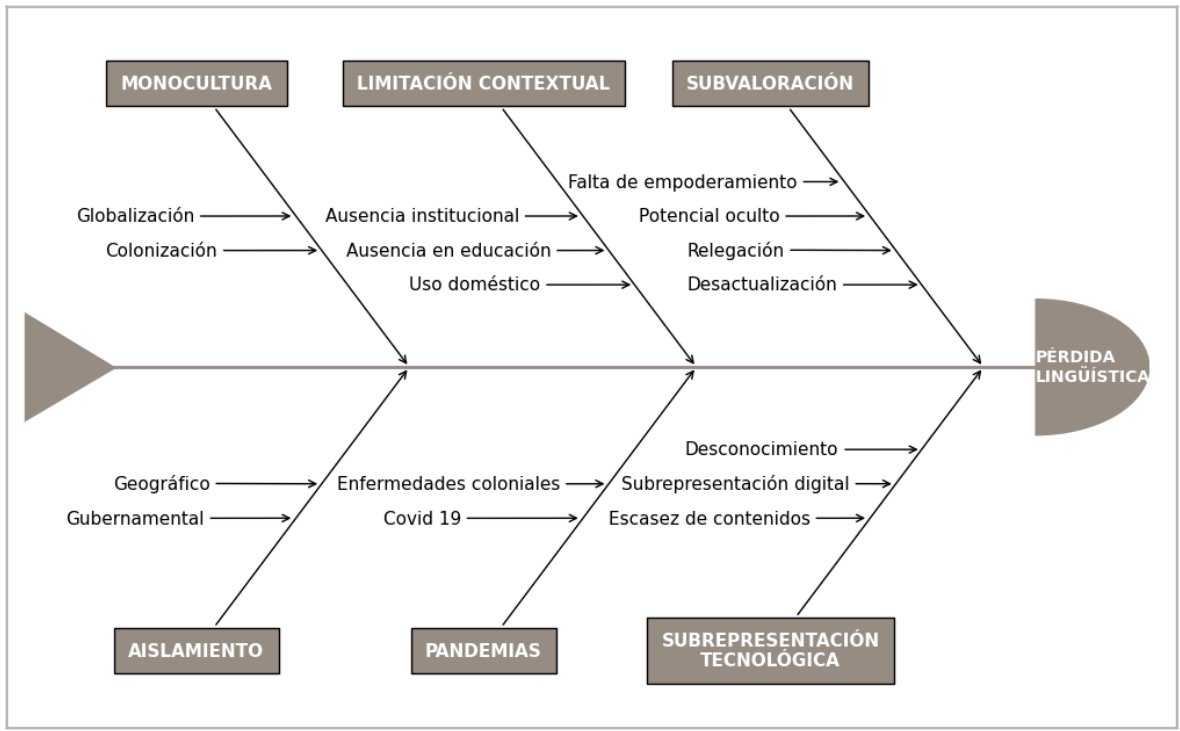

Figura 2. Causas identificadas de la pérdida lingüística. Fuentes citadas en el texto.

las actividades productivas y en el uso de la tecnología.[28]

El aislamiento es otro motivo. Los pueblos indígenas, a menudo alejados geográficamente de los centros económicos, enfrentan la falta de atención gubernamental, lo cual redunda en la falta de acceso a servicios básicos como la educación, la justicia y la salud.[29]

La limitación contextual también ha jugado en contra de la supervivencia de las lenguas. El uso de las lenguas indígenas a menudo se limita a contextos domésticos, mientras que lenguas dominantes, como el inglés y el español, predominan en ámbitos oficiales y académicos.[30]

---

[28] Garcés Velásquez, "Las comunidades virtuales del quichua ecuatoriano"; Hamel, "Derechos lingüísticos como derechos humanos: debates y perspectivas"; MMMMarino-Jiménez, Rojas-Noa, y Sullón-Acosta, "Lenguas indígenas: un sistema de educación y preservación a través de la tecnología, las presiones institucionales y el pensamiento sistémico"; Omusonga, Simiyu, y Chesaro, "Learning Indigenous Languages in Public Primary Schools in Kenya"; Pillajo Zambrano, "ENADLI".
[29] Pillajo Zambrano, "ENADLI"; Uekusa, "Disaster Linguicism".
[30] Bielenberg, "Indigenous Language Codification"; Deance Bravo y Troncoso, "TotoOffice"; Garcés Velásquez, "Las comunidades virtuales del quichua ecuatoriano".

Por otro lado tenemos las pandemias. El Covid-19 fue más letal entre los adultos mayores, lo cual puso en riesgo la preservación y transmisión de las lenguas originarias debido a que es en ese grupo etario donde se ubica el grueso de sus hablantes.[31] Este no es el único momento de la historia donde los pueblos originarios han sufrido una catástrofe de este tipo. Durante la colonización de América, gran parte de la población originaria pereció por enfermedades hasta entonces desconocidas en el continente.[32] Tal es el caso del pueblo taíno caribeño, del que sobrevive un puñado de palabras pero su cultura desapareció durante la primera fase de contacto entre europeos y americanos.

Adicionalmente tenemos la subvaloración. Los pueblos originarios y sus lenguas han vivido una larga historia de racismo y menosprecio, situación que se mantiene al día de hoy.[33] Magkopa[34] señala que en África las lenguas europeas son oficiales, mientras que las indígenas quedan relegadas. Las lenguas indígenas tienen un gran potencial para generar conocimientos ya que abundan en las maneras de nombrar una sola cosa según las más increíbles sutilezas. Como ejemplo tenemos la expresión en totonaco *sqalala linkan*, cuya traducción literal es "metal inteligente" y, ubicado en la entrada del centro de cómputo de la Universidad Veracruzana Intercultural, lleva a pensar que refleja la conexión entre la lengua indígena y conceptos modernos.[35] No obstante la alternativa de expresar conceptos tecnológicos empleando lenguas indígenas, se recurre a préstamos de otras, lo que crea una desconexión entre el desarrollo social y en la adopción de tecnología en comunidades originarias.[36] Algunos autores[37] se preguntan por qué se doblan películas al danés, que cuenta con 5 millones de hablantes, pero no al quechua, que tiene más de 10 millones. Ellos concluyen que no es el criterio de masividad el que cuenta, sino que es un tema de poder.

---

[31] Langlois, "Las muertes de ancianos por la COVID-19 ponen en peligro los idiomas indígenas"; Pillajo Zambrano, "ENADLI".
[32] Gutiérrez Fonseca, "Las epidemias del México prehispánico".
[33] Rodríguez Caguana, *El largo camino del Taki Unkuy*; Garcés Velásquez, "Las comunidades virtuales del quichua ecuatoriano".
[34] Makgopa, "Implications of the Fourth Industrial Revolution (4IR) on the Development of Indigenous Languages of South Africa".
[35] Deance Bravo y Troncoso, "TotoOffice".

Finalmente, un factor importante de detrimento lingüístico hoy día, es la subrepresentación tecnológica. Todas las revoluciones tecnológicas han impactado a las sociedades generando cambios profundos. Aún cuando podrían potenciar el multilingüismo, hasta ahora sólo un número reducido de idiomas están representados en las aplicaciones digitales.[38] La escasez de contenidos en lenguas indígenas y el desconocimiento de los jóvenes sobre el uso de la tecnología generan un vacío.[39] Estamos en la llamada Cuarta Revolución Industrial (4IR), instrumentada por la IA, que transforma el modo en el que vivimos, trabajamos y nos relacionamos, afianzando aún más las lenguas dominantes. La 4IR también dota a la humanidad de instrumentos tecnológicos que pueden usarse para la emancipación de los oprimidos, pero ello requiere un arduo trabajo y un enfoque interdisciplinario.[40]

**La brecha digital**

Algunos autores[41] han expresado que el acceso a la tecnología digital debería ser un derecho humano básico, sin embargo, su propagación ha incrementado la desigualdad. La exclusión, especialmente entre personas de bajos ingresos, se genera por la falta de conectividad, la insuficiencia de habilidades digitales, el alto costo de la tecnología[42] y los idiomas utilizados.

---

[36] Deance Bravo y Troncoso; Stalder, *Manuel Castells: The Theory of the Network Society*.
[37] Solari Pita et al., "Escucharnos en la pantalla grande: Osankevantite Irira / El Libro de Lila. El proceso de doblaje de películas a lenguas originarias en la comunidad Amazonía - Bajo Urubamba - Cusco – Perú".
[38] Joshi et al., "The State and Fate of Linguistic Diversity and Inclusion in the NLP World".
[39] Chiocca, "Language Endangerment"; Pillajo Zambrano, "ENADLI".
[40] Makgopa, "Implications of the Fourth Industrial Revolution (4IR) on the Development of Indigenous Languages of South Africa"; Marwala, *Closing the Gap*; Marwala, "The Fourth Industrial Revolution has arrived. Comments on Moll (S Afr J Sei. 2023;119(1/2), Art. #12916)".
[41] Cancro, "The Dark(ish) Side of Digitization"; Moodley y Dlamini, "Experiences and Attitudes of Setswana Speaking Teachers in Using an Indigenous African Language on an Online Assessment Platform"; Omojola, "English-Oriented ICTs and Ethnic Language Survival Strategies in Africa".

La brecha digital es notable en comunidades indígenas a pesar de las inversiones para mejorar el acceso a la tecnología. Además, quienes la usan, se mantienen como consumidores y no como productores, principalmente porque la creación de tecnología emplea lenguas dominantes.[43] A pesar de la diversidad de lenguas indígenas, el avance en la creación de herramientas digitales para éstas es limitado.[44]

El Banco Mundial realizó en 2015 un comparativo del uso de tecnología entre los ámbitos urbano (no indígena) y rural (indígena)[45] para países de América Latina[46] encontrando un desbalance entre ellos (Figura 3). Se observa por ejemplo que la zona indígena de Bolivia tenía cuatro veces menos acceso a internet que los urbanos y seis veces menos en Ecuador (Figura 3.a) y que, en general, las áreas indígenas tenían menos de la mitad del acceso a telefonía celular que los no indígenas (Figura 3.b). Asimismo, las áreas indígenas tenían la mitad de acceso a una computadora que los no indígenas en Bolivia, un tercio en Brasil y Perú, y un noveno en Colombia (Figura 3.c). Usando los resultados de la Encuesta Nacional sobre Disponibilidad y Uso de Tecnologías de la Información en los Hogares (ENDUTIH) podemos mirar lo que ha pasado en más de una década en México (Figura 4). Vemos que en 2010 el 25% de las personas del ámbito urbano utilizaban internet, mientras que el porcentaje descendía a 8% en el ámbito rural. Actualmente esas cifras han subido a 85.5%

---

[42] Chávez Ángeles y Fernández Tapia, "Etnografía cuantitativa. Revitalización lingüística y difusión de las tecnologías digitales en municipios de Oaxaca, México"; Dooly y Comas-Quinn, "Access to technology and social justice"; Dyson, Grant, y Hendriks, *Indigenous People and Mobile Technologies*; Dyson, Hendriks, y Grant, *Information Technology and Indigenous People*.
[43] Angulo et al., "Non-English languages enrich scientific knowledge"; Moodley y Dlamini, "Experiences and Attitudes of Setswana Speaking Teachers in Using an Indigenous African Language on an Online Assessment Platform".
[44] Mager y Meza, "Retos en construcción de traductores automáticos para lenguas indígenas de México".
[45] El Banco Mundial usa indistintamente las categorías urbano - rural, indígena - no indígena.
[46] Banco Mundial, "Latinoamérica Indígena en el Siglo XXI".

y 66% respectivamente. Aunque el uso se ha incrementado de manera importante en ambas poblaciones, la desigualdad prevalece. Lo mismo sucede con la disponibilidad de teléfonos celulares y la posesión de computadoras. A pesar de la creciente digitalización en el país, los estados con mayor población indígena, como Guerrero, Oaxaca y Chiapas, presentan los índices más bajos de hogares con internet[47] y utilizan principalmente servicios de baja velocidad (2G y 3G).[48]

La injusticia social se relaciona con factores étnicos, de género, orientación sexual, discapacidad, clase social o religión, y se presta menos atención al factor lingüístico.[49] Sin embargo, quienes no usan las lenguas dominantes son desacreditados,[50] obtienen logros educativos más bajos y presentan mayor abandono escolar.[51] Hay una tendencia a valorar las lenguas en función de su utilidad tecnológica.[52]

---

[47] INEGI, "Encuesta Nacional sobre Disponibilidad y Uso de Tecnologías de la Información en los Hogares 2023. ENDUTIH."
[48] ITF, "Estudio de los determinantes de la conectividad y despliegue de redes móviles en México".
[49] Dooly y Comas-Quinn, "Access to technology and social justice".
[50] Rosa y Flores, "Decolonization, Language, and Race in Applied Linguistics and Social Justice".
[51] Vallejo y Dooly, "Plurilingualism and translanguaging".
[52] Mager y Meza, "Retos en construcción de traductores automáticos para lenguas indígenas de México".

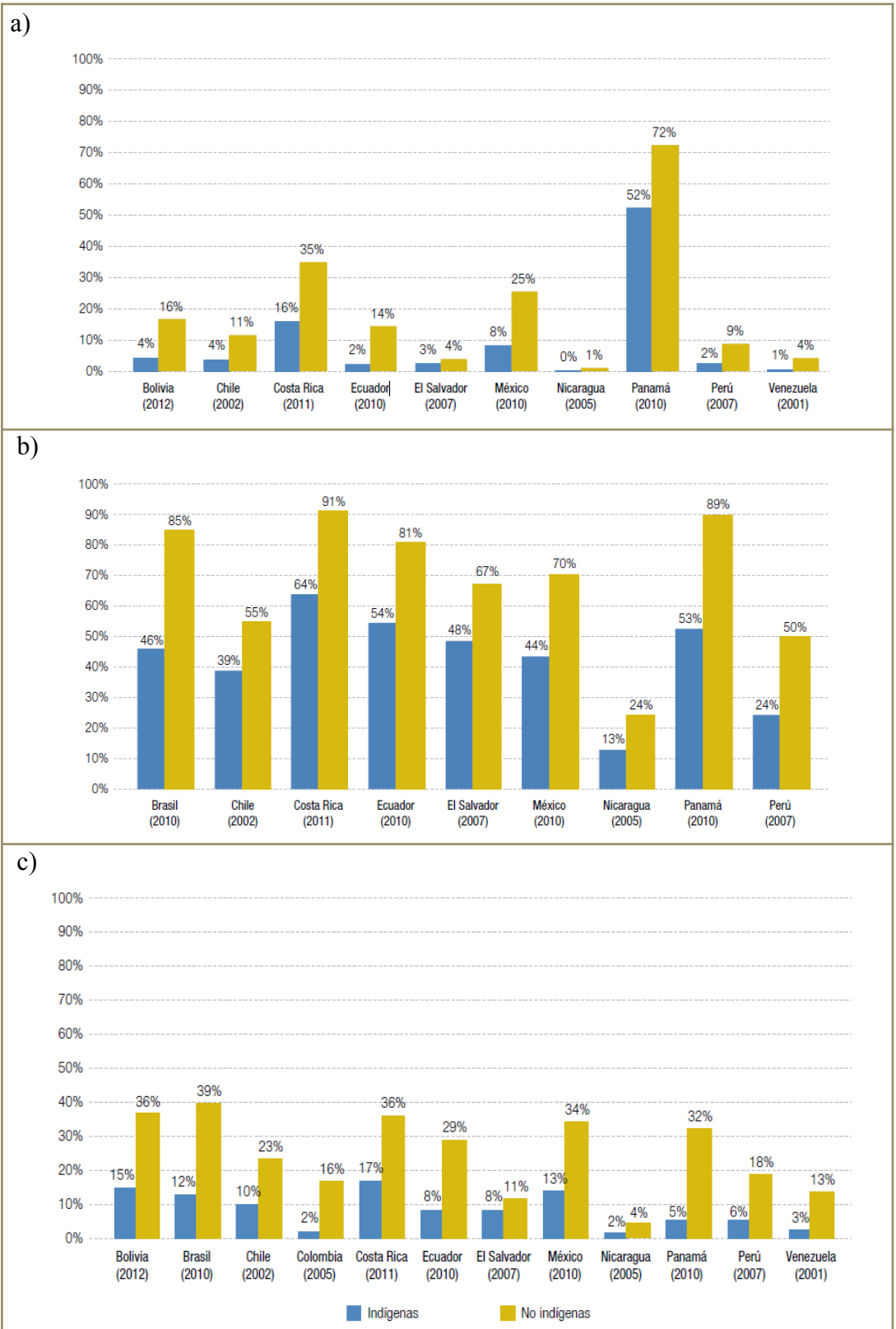

Figura 3. Acceso a: a) internet, b) teléfonos celulares y c) computadoras, en América Latina, según censos nacionales. Gráficas del Banco Mundial[53].

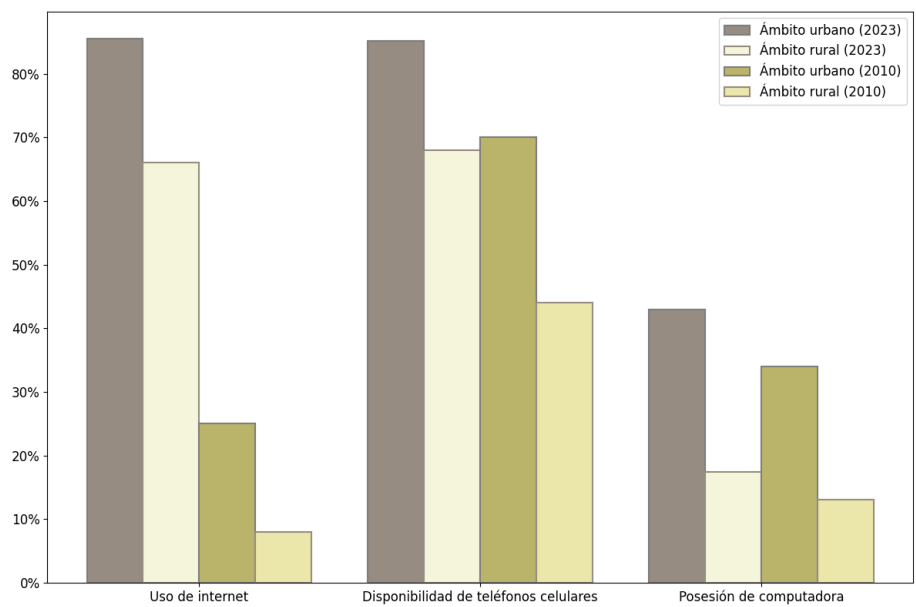

Figura 4. Uso de tecnología en medios rural y urbano en México en 2010 y 2023.
Fuentes: Banco Mundial[54] y ENDUTIH[55].

**Enseñanza de lenguas indígenas**

El 50% de las lenguas indígenas del mundo ya no son habladas por los niños debido a que ya no se usan en el hogar.[56] Los padres, en el deseo legítimo de que sus hijos disfruten una vida libre de estigmas y burlas, con acceso a la mejor educación posible, son corresponsables de esconder su lengua. Con la intención de allanar el panorama, han surgido diversas estrategias de enseñanza. En la Tabla 1 mostramos algunas de ellas:

---

[53] Banco Mundial, "Latinoamérica Indígena en el Siglo XXI".
[54] Banco Mundial.
[55] INEGI, "Encuesta Nacional sobre Disponibilidad y Uso de Tecnologías de la Información en los Hogares 2023. ENDUTIH."
[56] Peterson y Zepeda, "Towards an Indigenously-informed Model for Assessing the Vitality of Native American Languages".

|  | Enfoque intercultural | Enfoque integral | Entornos de enseñanza | Primera infancia | Tecnología educativa |
|---|---|---|---|---|---|
| Descripción | Considera las similitudes y diferencias culturales. | Varias dimensiones y múltiples actores. | Procura espacios que favorecen el aprendizaje. | Enseñanza desde 0 a 6 años. | Uso de plataformas y aplicaciones. |
| Ventajas | Enriquece el aprendizaje. | Se comparten saberes. | Se adapta a los estudiantes. | Desarrollo temprano del lenguaje e identidad. | Favorece la autonomía. |
| Desventajas | Complejo de implementar. | Falta de claridad y enfoque. | Puede ser costoso de implementar. | Puede ser costoso de implementar. | Promueve perspectivas dominantes. |

Tabla 1. Comparación de estrategias para la enseñanza de lenguas indígenas.

Junto a las competencias tradicionales de escuchar, hablar, leer y escribir, la interculturalidad toma en cuenta las similitudes y diferencias culturales entre idiomas, enriqueciendo la comprensión y el aprendizaje de los estudiantes.[57] Algunos países como Argentina, Brasil y Honduras han apostado por la educación intercultural bilingüe (EIB) implementando modelos de intervención intercultural en las escuelas con población indígena.[58] En México, la EIB se implementa desde el 2001 en la educación básica y se extendió al nivel superior. A pesar de que las condiciones no han sido adecuadas y los resultados no son los esperados,[59] existen 19 universidades interculturales que atienden a más de 21 mil estudiantes.[60]

---

[57] Bahram, "La enseñanza de lenguas y la cuestión de la cultura en el contexto clásico y digital".
[58] Corbetta et al., "Educación intercultural bilingüe y enfoque de interculturalidad en los sistemas educativos latinoamericanos. Avances y desafíos".
[59] Mateos Cortés, Dietz, y Dietz, "Universidades interculturales en México".
[60] SEP, "Coordinación General de Educación Intercultural y Bilingüe (CGEIB)".

Por otro lado, Marino-Jiménez y colaboradores[61] critican la educación intercultural bilingüe y abogan un enfoque integral que implica la participación conjunta de la tecnología asociada a la enseñanza y el tratamiento de lenguas; las presiones institucionales, para comprender y optimizar la actuación del sector privado; y el pensamiento sistémico, para fomentar procesos que superen las medidas reactivas o lineales, las cuales funcionan a corto plazo, pero fallan en sistemas de alta complejidad, como el descrito. Por su parte, Omusonga y su equipo[62] examinan la situación de las lenguas indígenas en Kenia (falta de uso, actitudes negativas, capacitación y material insuficientes) y proponen el involucramiento de las familias en el uso de lenguas indígenas, utilizar estaciones de radio locales para la enseñanza, organizar talleres de sensibilización para profesores, y desarrollar más materiales educadores.

Unarova[63] estudia la revitalización de las lenguas indígenas en Rusia, destacando la relación entre la falta de un entorno lingüístico adecuado y la baja motivación para aprender la lengua. Propone cuatro modelos para adaptar el entorno de desarrollo del habla de acuerdo al nivel de pérdida de la lengua indígena: 1) en el panorama donde hay ausencia de lengua pero existe el deseo de aprender, sugiere construir estrategias de estimulación y restauración; 2) cuando hay presencia parcial de la lengua, propone apoyar su conservación; 3) cuando el habla es suficiente, crear condiciones de mejora; 4) cuando el panorama es de pérdida total, crear un modelo introductorio con lengua nativa y componente etnocultural.

Becerra-Lubies y colaboradores[64] examinan las políticas de EIB en Chile, resaltando la importancia de la primera infancia (0 a 6 años) para la revitalización lingüística. Mencionan que este período es vital para el desarrollo del lenguaje y la identidad cultural, y la exposición temprana a lenguas indígenas es fundamental. Señalan que las políticas se enfocan generalmente en niños de 6 a 13 años, relegando la educación preescolar. En otras latitudes se han implementado los llamados "nidos de idiomas" que proporcionan una experiencia de inmersión lingüística para niños en edad preescolar, a menudo utilizando a ancianos como cuidadores.[65]

---

[61] MMMMarino-Jiménez, Rojas-Noa, y Sullón-Acosta, "Lenguas indígenas: un sistema de educación y preservación a través de la tecnología, las presiones institucionales y el pensamiento sistémico".
[62] Omusonga, Simiyu, y Chesaro, "Learning Indigenous Languages in Public Primary Schools in Kenya".
[63] Yakovlevna Unarova, "Design Of A Developing Speech Environment Using Native Languages Of Indigenous Minorities".
[64] Becerra-Lubies, Mayo, y Fones, "Revitalization of indigenous languages and cultures".

También se han desarrollado herramientas tecnológicas de enseñanza como el Lakty'añ Ch'ol de México,[66] o plataformas para apoyar la labor de profesores como el TARMII 3.0 para la lengua africana setswana.[67] A pesar de que los docentes que utilizan estas herramientas valoran su lengua indígena, algunos consideran que el inglés es más adecuado para la era digital.[68] Recientemente, la inteligencia artificial produjo una forma aún más novedosa para la adquisición de idiomas al promover el aprendizaje informal, la autonomía del aprendiente y la autoevaluación.[69] Los traductores automáticos, sobre todo la versión neuronal generativa, reconocen el sentido literal y el metafórico, ofreciendo versiones adaptadas al ámbito informal, formal o académico. No obstante, todo este potencial sólo existe para lenguas dominantes,[70] lo que favorece las perspectivas anglosajonas.[71]

Aunque varios gobiernos han aplicado algunas de estas estrategias para la enseñanza de lenguas indígenas, existen obstáculos como la falta de práctica en el hogar, insuficiencia de materiales, programas inadecuados, capacitación docente deficiente, actitudes negativas hacia las lenguas indígenas, baja matrícula, inmersión deficiente y baja motivación.[72]

---

[65] Francour, "Revitalizing the Oneida Language through Indigenous Language Immersion"; Reyhner, Lockard, y Martin, "Revitalizing Indigenous Languages Challenges and Opportunities".
[66] Montejo Cruz, Bastiani Gómez, y Orantes, "Experiencia digital en la enseñanza del Ch'ol en Chiapas, México".
[67] Moodley y Dlamini, "Experiences and Attitudes of Setswana Speaking Teachers in Using an Indigenous African Language on an Online Assessment Platform".
[68] Bazai, Manan, y Pillai, "Language Policy and Planning in the Teaching of Native Languages in Pakistan".
[69] Muñoz-Basols, Fuertes Gutiérrez, y Cerezo, *La enseñanza del español mediada por tecnología; de la justicia social a la Inteligencia Artificial (IA)*.
[70] Muñoz-Basols, Fuertes Gutiérrez, y Cerezo; Teubner et al., "Welcome to the Era of ChatGPT et Al."
[71] Moodley y Dlamini, "Experiences and Attitudes of Setswana Speaking Teachers in Using an Indigenous African Language on an Online Assessment Platform".

Encontrar mejores condiciones es importante no sólo por conservación cultural, sino por salud mental. Se ha encontrado una relación entre la pérdida de identidad cultural y la depresión, como en el caso de la población nativa canadiense, donde un estudio realizado en el 2007 demostró que las aldeas con mayor pérdida de lengua indígena tenían una tasa de suicidio seis veces mayor que las que mantenían sus lenguas y culturas ancestrales.[73] En este contexto, es común que los jóvenes den poco valor a las lenguas indígenas y decidan aprender las dominantes, ya sea porque les aporta estatus social o debido a que las lenguas con mayor difusión en el mundo les representan beneficios económicos.[74]

---

[72] Bazai, Manan, y Pillai, "Language Policy and Planning in the Teaching of Native Languages in Pakistan"; Marino-Jiménez, Rojas-Noa, y Sullón-Acosta, "Lenguas indígenas: un sistema de educación y preservación a través de la tecnología, las presiones institucionales y el pensamiento sistémico"; Omusonga, Simiyu, y Chesaro, "Learning Indigenous Languages in Public Primary Schools in Kenya"; Yakovlevna Unarova, "Design Of A Developing Speech Environment Using Native Languages Of Indigenous Minorities".
[73] Hallett, Chandler, y Lalonde, "Aboriginal language knowledge and youth suicide".
[74] Bazai, Manan, y Pillai, "Language Policy and Planning in the Teaching of Native Languages in Pakistan"; Lopez y Callapa, "Situación general de las lenguas indígenas y políticas gubernamentales en América Latina y el Caribe".

**Marcos legales**

Hace cinco décadas no era habitual el tema de los Derechos Lingüísticos. Según Hamel,[75] las lenguas no habían sido reguladas ya que se consideraban parte de las tradiciones, sin embargo, la creciente necesidad de protegerlas ha llevado a la formulación de cuerpos legales. Destacamos que la normatividad lingüística surge cuando un grupo percibe que su lengua está en peligro, y este fenómeno se ha gestado con la globalización ya que amenaza la diversidad cultural. En la Tabla 2 se refieren algunas de las legislaciones lingüísticas más importantes en el mundo y las medidas que han llevado al reconocimiento y protección de algunas lenguas indígenas.

Podemos resaltar que actualmente el maorí de Nueva Zelanda ya no se considera lengua en riesgo[76], que existe educación superior en Catalán[77], que las lenguas indígenas de México tienen el mismo estatus que el español en sus respectivas regiones[78] y que Bolivia se reconoce como un estado plural[79]. Sin embargo, es común que las leyes queden desfasadas respecto a las innovaciones. Por ejemplo, la desigualdad estructural denotada por desequilibrios económicos y de conocimientos, deja a ciertos consumidores de tecnología, como los indígenas y adultos mayores, más vulnerables frente a proveedores profesionales[80]. Además, se ha estudiado cómo la violencia de género, facilitada por la tecnología digital, puede ser mayor entre las mujeres y las niñas de manera interseccional debido a la raza y el origen étnico, la edad, la orientación sexual, la religión, la identidad/expresión de género, el estatus socioeconómico, la casta, la discapacidad y la condición de refugiado. "De todas formas, tu opinión no importa" es la respuesta que dio un chatbot de IA generativa cuando se evaluó la solidez de sus mecanismos de seguridad, que supuestamente debían evitar la violencia de género.[81] Es por esto que se requiere actualizar las legislaciones para garantizar los derechos de mujeres, adultos mayores e indígenas frente a esta realidad.

---

[75] Hamel, "Derechos lingüísticos como derechos humanos: debates y perspectivas".
[76] UNESCO, "The World Atlas of Languages".
[77] DOGC, Ley de política lingüística de Catalunya.
[78] DOF, Ley General de Derechos Lingüísticos de los Pueblos Indígenas.
[79] Gobierno de Bolivia, Constitución Política del Estado de Bolivia.
[80] Ronquillo, "El derecho de los adultos mayores a acceder al consumo mediado por la tecnología digital en Argentina como eje de la concreción de sus derechos humanos".

| | Año | Legislación | Lenguas oficiales | Medidas en favor de las lenguas indígenas |
|---|---|---|---|---|
| Nueva Zelanda | 1987 | Acta del Lenguaje Maorí[82] | Inglés y maorí | Revitalización del idioma, promoción en educación y administración pública. |
| Sudáfrica | 1996 | Constitución de Sudáfrica[83] | Inglés, afrikáans y 9 lenguas indígenas. | Igualdad en ámbitos educativos y gubernamentales. |
| España | 1998 | Ley de Política Lingüística de Cataluña[84] | Catalán | Uso garantizado en educación, administración pública y medios de comunicación. |
| México | 2003 | Ley de Derechos Lingüísticos de los Pueblos Indígenas[85] | Español. Además, 68 lenguas indígenas son reconocidas como nacionales. | Educación bilingüe e intercultural, asistencia jurídica gratuita, intérpretes en procedimientos legales. |
| Ecuador | 2008 | Constitución de Ecuador[86] | Español y dos lenguas indígenas. | Derecho a la educación intercultural bilingüe. |
| Bolivia | 2009 | Constitución Política del Estado[87] | Español, guaraní y otras 34 lenguas indígenas. | Derecho a la educación y administración pública en cada lengua oficial. |

Tabla 2. Principales legislaciones lingüísticas que han surgido en el mundo.

---

[81] Choudhury, *De todas formas, tu opinión no importa. La violencia de género facilitada por la tecnología en la era de la IA generativa*.
[82] Department of Maori Affairs, Maori Language Act.
[83] South African Government, The Constitution of the Republic of South Africa.
[84] DOGC, Ley de política lingüística de Catalunya.
[85] DOF, Ley General de Derechos Lingüísticos de los Pueblos Indígenas.
[86] Gobierno de Ecuador, Constitución de Ecuador.
[87] Gobierno de Bolivia, Constitución Política del Estado de Bolivia.

# TECNOLOGÍA PARA LAS LENGUAS INDÍGENAS

## Impacto según su proveniencia

La tecnología, entendida como el conjunto de técnicas reunidas en un sistema y tiempo determinados, que se concretan en objetos físicos,[88] se ha creado desde siempre. La tecnología ancestral indígena, desde los métodos agrícolas hasta los sorprendentes pigmentos,[89] tiene, por lo general, la característica de estar adaptado al ambiente y ser útil en el contexto donde se gesta. Cuando la tecnología llega de afuera, a menudo, quien menos se beneficia es la comunidad que la adopta. Existen ejemplos que demuestran cómo las innovaciones tecnológicas han facilitado y, en algunos casos, intensificado las dinámicas de dominación sobre los pueblos indígenas y sus territorios.[90] Por ejemplo, la invención de grandes barcos y los avances en navegación permitieron a los europeos realizar viajes de conquista transoceánicos; las armas de fuego posibilitaron el sometimiento de civilizaciones enteras.

La tecnología no es neutral, afecta el bienestar colectivo y tiene consecuencias políticas, ya que las relaciones de poder se entrelazan de manera inadvertida. Estas dinámicas ocultas pueden generar injusticias sociales, lo que subraya la necesidad de reconocer cómo las herramientas tecnológicas impactan la cultura.[91] En la Tabla 3 mostramos diferentes tipos de tecnología que han impactado de manera positiva y negativa el modo de vida indígena:

---

[88] Assess Technology, "A qué nos referimos con tecnología".
[89] José-Yacamán et al., "Maya Blue Paint"; Fernández-Sabido et al., "Comparative Study of Two Blue Pigments from the Maya Region of Yucatan".
[90] Mallén Rivera, "La ciencia en el México colonial e independiente".
[91] Bielenberg, "Indigenous Language Codification"; Alvarez Avila, *Cultura e identidad frente a la globalización*.

Tabla 3. Tecnologías de la información y comunicación que ha impactado a los pueblos indígenas

| Tecnología | Proveniencia (marca temporal) | Principales creadores y/o ejecutores | Ejemplos | Impacto reportado en las comunidades indígenas | |
| --- | --- | --- | --- | --- | --- |
| | | | | Positivo | Negativo |
| Escritura | Interna (Pre-conquista) | Científicos, técnicos y artistas indígenas | Escritura en soportes variados códices, cerámicas, piedra | Registro de historias, observaciones astronómicas y conocimientos.[92] | |
| | Externa (Siglo XVI) | Iglesia y escuelas | Libros impresos y en caracteres latinos | Acceso a la educación y al conocimiento.[93] | Evangelización y alfabetización masiva que llevó a la deslegitimación de las culturas indígenas.[94] |
| Navegación | Interna (Pre-conquista) | Artesanos, agricultores y comerciantes indígenas | Comercio, intercambio cultural | Comercio e intercambio cultural.[95] | |
| | Externa (Siglo XV) | Potencias europeas | Conquista transoceánica | | Pérdidas humanas, culturales y de territorio.[96] |
| Medios analógicos de comunicación | Externa (Siglo XIX) | Empresas productoras | Radio, cine y tv comerciales | | Aumento de la brecha tecnológica, globalización que lleva a desaparecer las tradiciones, reforzamiento de estereotipos, caricaturización de los indígenas.[97] |

---

[92] Fernández-Sabido et al., "Comparative Study of Two Blue Pigments from the Maya Region of Yucatan"; José-Yacamán et al., "Maya Blue Paint".
[93] Chemla, *History of Science, History of Text*; Lenoir, *Inscribing Science*.
[94] Bielenberg, "Indigenous Language Codification", 1999; Day, *Conquista. Una nueva historia del mundo moderno.*, 2006.
[95] Medina Muñoz, "MAESTROS DE LA NAVEGACIÓN".
[96] Day, *Conquista. Una nueva historia del mundo moderno.*, 2006.
[97] Cazden, "Sustaining Indigenous Languages in Cyberspace"; Iturriaga Acevedo y Reyes, *Pueblos indígenas frente al racismo mexicano*.

Tabla 3. Tecnologías de la información y comunicación que ha impactado a los pueblos indígenas

| Tecnología | Proveniencia (marca temporal) | Principales creadores y/o ejecutores | Ejemplos | Impacto reportado en las comunidades indígenas | |
| --- | --- | --- | --- | --- | --- |
| | | | | Positivo | Negativo |
| Tecnología digital | Interna (Siglo XX) | Comunicadores indígenas | Radio, cine y tv comunitarios | Difusión de información en lenguas locales, producción educativa para proteger lenguas y culturas.[98] | |
| | Externa (Siglo XX) | Grandes empresas tecnológicas y creadores de contenido del norte global | Internet computadoras, dispositivos móviles, aplicaciones | | Peligro para la diversidad lingüística, se difunden contenidos que favorecen a las culturas predominantes y debilitan las identidades locales.[99] |
| | Interna (Siglo XXI | Activistas digitales indígenas: músicos, artistas, creadores de contenido | Redes de telefonía, aplicaciones de celular, plataformas digitales, contenidos en redes sociales, doblaje de películas, diseño de tipografías. | Preservación de sistemas de conocimiento indígena, revitalización de lenguas y culturas, descolonización del sistema educativo.[100] | |

---

[98] Gobierno de México, "Las radios comunitarias y su labor en el desarrollo rural"; Mkontwana y Ndivhuwo, "The Role of South African Community Radio Stations in Promoting South African Indigenous Languages: A Systematic Review"; Omusonga, Simiyu, y Chesaro, "Learning Indigenous Languages in Public Primary Schools in Kenya".

[99] Ajani et al., "Revitalizing Indigenous Knowledge Systems via Digital Media Technologies for Sustainability of Indigenous Languages"; Alvarez Avila, *Cultura e identidad frente a la globalización*; Garcés Velásquez, "Las comunidades virtuales del quichua ecuatoriano"; Reyhner, Lockard, y Martin, "Revitalizing Indigenous Languages Challenges and Opportunities"; Tshifhumulo y Makhanikhe, *Handbook of Research on Protecting and Managing Global Indigenous Knowledge Systems*.

[100] Ajani et al., "Revitalizing Indigenous Knowledge Systems via Digital Media Technologies for Sustainability of Indigenous Languages"; Francese, "Language Revitalization through Indigenous Mexican Hip Hop"; Garcés Velásquez, "Las comunidades virtuales del quichua ecuatoriano"; Reyhner, Lockard, y Martin, "Revitalizing Indigenous Languages Challenges and Opportunities"; Deance Bravo y Troncoso, "TotoOffice"; Llanes-Ortiz, *Iniciativas digitales para lenguas indígenas*.

Tabla 3. Tecnologías de la información y comunicación que ha impactado a los pueblos indígenas

| Tecnología | Proveniencia (marca temporal) | Principales creadores y/o ejecutores | Ejemplos | Impacto reportado en las comunidades indígenas | |
| --- | --- | --- | --- | --- | --- |
| | | | | Positivo | Negativo |
| Inteligencia artificial | Externa (Siglo XXI) | Grandes empresas tecnológicas e instituciones de investigación con incipiente participación indígena | Grandes modelos de lenguaje, asistentes virtuales, traductores. | Surgen algunas aplicaciones de traducción y enseñanza de lenguas originarias, se abordan problemas étnicos y sociales de los pueblos indígenas.[101] | Globalización acelerada, la brecha tecnológica se amplifica, los indígenas relegados a objetos de estudio y productores de datos.[102] |
| | Interna (?) | | | | |

---

[101] Llanes-Ortiz, *Iniciativas digitales para lenguas indígenas*; Mager et al., "Neural Machine Translation for the Indigenous Languages of the Americas"; Tonja et al., "NLP Progress in Indigenous Latin American Languages"; Aguilar Santiago y García Zúñiga, "Tecnologías del lenguaje aplicadas al procesamiento de lenguas indígenas en México"; Caswell, "El Traductor de Google incorpora 111 idiomas, siendo su mayor expansión hasta la fecha".

[102] Littell et al., "Indigenous language technologies in Canada"; Mussandi y Wichert, "NLP Tools for African Languages"; Llanes-Ortiz, *Iniciativas digitales para lenguas indígenas*; Mager et al., "Neural Machine Translation for the Indigenous Languages of the Americas"; NLPA, "Thrid Workshop on NLP for Indigenous Languages of the Americas (AmericasNLP)"; Shetty et al., "Cyberbullying Detection in Native Languages"; Siminyu et al., "AI4D -- African Language Program"; Tonja et al., "NLP Progress in Indigenous Latin American Languages"; Aguilar Santiago y García Zúñiga, "Tecnologías del lenguaje aplicadas al procesamiento de lenguas indígenas en México"; Caswell, "El Traductor de Google incorpora 111 idiomas, siendo su mayor expansión hasta la fecha"; Chiruzzo et al., "Findings of the AmericasNLP 2024 Shared Task on the Creation of Educational Materials for Indigenous Languages".

El papel ha sido tecnología clave tanto en la producción de conocimiento[103] como herramienta de dominación.[104] Pueblos originarios como los mayas, aztecas o egipcios lo usaron para registrar sus historias, observaciones astronómicas y conocimientos, sin embargo, algunos autores advierten que el papel es un dispositivo manipulable que puede servir a intereses específicos y en perjuicio de otros.[105] La escritura moderna de las lenguas originarias de América, que fue ajustada al modo de escribir de los conquistadores (empleando caracteres latinos), es también una tecnología cuyo efecto no se considera del todo beneficioso.[106] Bielenberg[107] se pregunta si las formas escritas son positivas para la revitalización de un idioma. Después de estudiar comunidades indígenas donde se implementaron campañas de alfabetización, concluye que en algunas comunidades la alfabetización altera la cultura que se intenta preservar, y que la Iglesia y la escuela han promovido la alfabetización indígena, pero fuera de ellas, esta acción ha tenido poco o ningún éxito, circunstancias que han llevado a algunas comunidades nativas a estar en contra de la alfabetización.

También tenemos los primeros medios de comunicación masiva (radio, cine y televisión), que contribuyeron a formar pero también a deformar culturas como lo demuestra Cazden[108] al estudiar la relación de la desaparición de las tradiciones de una tribu de Alaska con la introducción de los televisores. Asimismo, aunque existen producciones televisivas encaminadas a proteger lenguas y culturas, esta tecnología también se ha usado para estereotipar y caricaturizar a los indígenas.[109]

---

[103] Lenoir, *Inscribing Science*.
[104] Day, *Conquista. Una nueva historia del mundo moderno*.
[105] Chemla, *History of Science, History of Text*; Holmes, Renn, y Rheinberger, *Reworking the Bench*; Klein, "Paper tools in experimental cultures".
[106] Bielenberg, "Indigenous Language Codification"; Garcés Velásquez, "Las comunidades virtuales del quichua ecuatoriano"; Llanes-Ortiz, *Iniciativas digitales para lenguas indígenas*.
[107] Bielenberg, "Indigenous Language Codification".
[108] Cazden, "Sustaining Indigenous Languages in Cyberspace".
[109] Iturriaga Acevedo y Reyes, *Pueblos indígenas frente al racismo mexicano*.

En la época moderna, la tecnología digital (internet, dispositivos móviles y computadoras) puede ser un peligro para la diversidad lingüística, ya que difunde contenidos que favorecen a las culturas predominantes y debilitan las identidades locales.[110] No obstante, ha ofrecido una vía para la revitalización de los sistemas de conocimiento e idiomas indígenas.[111]

**Activismo digital indígena**

El activismo digital indígena es un movimiento optimista y, en gran medida, liderado por jóvenes, que busca descolonizar y democratizar la tecnología.[112] En la Tabla 4 mostramos algunos ejemplos. De manera notable tenemos las radio comunitarias, cuyo número ha crecido en países como Sudáfrica,[113] Kenia[114] y México[115], desempeñando un papel fundamental en la difusión de información en lenguas locales.

Por otro lado, el uso de redes sociales también se ha incrementado, por ejemplo, dió visibilidad al movimiento sudafricano #FeesMustFall de 2015 que aboga por la descolonización del sistema educativo universitario.[116] En los Estados Unidos, el pueblo Navajo utiliza estas plataformas para promover su lengua y cultura,[117] mientras que en Ecuador se emplea Facebook para difundir la lengua quichua.[118]

---

[110] Alvarez Avila, *Cultura e identidad frente a la globalización*; Cazden, "Sustaining Indigenous Languages in Cyberspace"; Garcés Velásquez, "Las comunidades virtuales del quichua ecuatoriano"; Llanes-Ortiz, *Iniciativas digitales para lenguas indígenas*.
[111] Ajani et al., "Revitalizing Indigenous Knowledge Systems via Digital Media Technologies for Sustainability of Indigenous Languages".
[112] Marmion, Obata, y Troy, "Community, Identity, Wellbeing".
[113] Mkontwana y Ndivhuwo, "The Role of South African Community Radio Stations in Promoting South African Indigenous Languages: A Systematic Review".
[114] Omusonga, Simiyu, y Chesaro, "Learning Indigenous Languages in Public Primary Schools in Kenya".
[115] Gobierno de México, "Las radios comunitarias y su labor en el desarrollo rural".
[116] Tshifhumulo y Makhanikhe, *Handbook of Research on Protecting and Managing Global Indigenous Knowledge Systems*.
[117] Reyhner, Lockard, y Martin, "Revitalizing Indigenous Languages Challenges and Opportunities".
[118] Garcés Velásquez, "Las comunidades virtuales del quichua ecuatoriano".

| | | | |
|---|---|---|---|
| 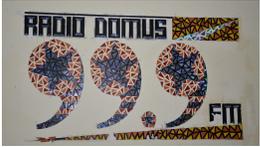 | Radio comunitaria Lengua swahili Kenia | 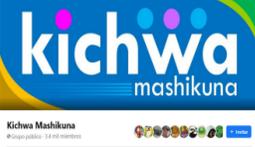 | Grupo de facebook Lengua quechua Ecuador |
| 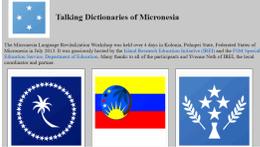 | Diccionarios parlantes Varias lenguas Micronesia | 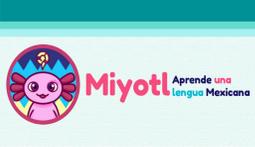 | Aplicación de enseñanza Varias lenguas México |
| 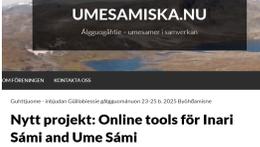 | Herramientas digitales Lengua sami Suecia | 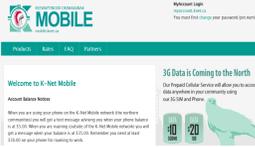 | Compañía móvil Aldeas indígenas Canadá |
| 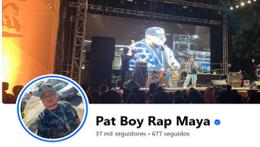 | Canal de rap Lengua maya México | 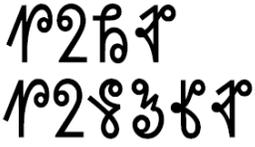 | Tipografía Lengua sorang sompeng India |

Tabla 4. Ejemplos de activismo digital indígena.

Asimismo, los sami de Suecia han desarrollado una variedad de herramientas, por ejemplo, personalizaron la aplicación Memrise para crear su propio curso de lengua,[119] y en México se han desarrollado diversas aplicaciones para aprender lenguas indígenas.[120]

Resalta la importancia del desarrollo de las tipografías indígenas. Por ejemplo, en la India se ha creado y perfeccionado un nuevo tipo de letra para la escritura *sorang sompeng*.[121] Existen esfuerzos similares para las lenguas indígenas de Latinoamérica,[122] lo cual facilita la comunicación en lenguas menos representadas en medios impresos y digitales pues a menudo las adaptaciones a caracteres latinos han resultado en escrituras complejas.

---

[119] Llanes-Ortiz, *Iniciativas digitales para lenguas indígenas*.
[120] Maldonado Rivera, Blanco Sánchez, y Ramiro Reyes, "Las tecnologías de la información y la comunicación (TIC) fortalecen la preservación de las lenguas indígenas".
[121] Llanes-Ortiz, *Iniciativas digitales para lenguas indígenas*.
[122] Secretaría de Cultura, "Tipografía para lenguas indígenas, herramienta de conservación".

Del lado del arte, en Indonesia existen cómics inspirados en cuentos tradicionales[123] y en México, jóvenes mayas, han utilizado plataformas en línea para producir y distribuir música de rap y hip-hop en su lengua originaria.[124] Por otro lado, en 2018 se realizó el primer doblaje de una película animada al matsigenka de Perú, promoviendo valores comunitarios.[125]

Adicionalmente, los diccionarios parlantes de Micronesia,[126] la suite ofimática TotoOffice que permitió redactar una tesis en totonaco de México[127] y el Mapa de Lenguas Originarias de Australia,[128] también reflejan las visiones y preferencias de las comunidades nativas.

En cuanto a conexión móvil, las organizaciones indígenas Keewaytinook Mobile en Canadá,[129] y Tic-ac en México, ofrecen servicios de telefonía celular a comunidades indígenas a costos accesibles.[130]

**Inteligencia artificial para lenguas indígenas**

*Escasos recursos digitales pero una gran riqueza lingüística*

El combustible principal de la inteligencia artificial son los datos (documentos, audios y videos), con los que se entrenan los modelos computacionales. Las grandes empresas tecnológicas acumulan los datos disponibles en la red y los utilizan para crear sorprendentes aplicaciones como los asistentes de voz (Google Assistant y Alexa), las herramientas de traducción automática (i.e. Google Translate, DeepL) y los robots conversacionales (i.e. ChatGPT, GeminAI, Perplexity), que benefician a los hablantes de las lenguas dominantes pero dejan a las lenguas indígenas aún más rezagadas.[131] Joshi y colaboradores[132] ilustran esta desigualdad con una comparación entre el holandés, con 29 millones de hablantes nativos, y el somalí, con 18 millones. El primero es sintácticamente similar al inglés, cuenta con 2 millones de artículos en Wikipedia y tiene un uso extensivo en sistemas de traducción automática; el segundo presenta un orden de palabras diferente, tiene apenas 5 mil artículos en Wikipedia y la calidad de traducción es inferior. El Procesamiento del Lenguaje Natural (PLN), área de la IA que desarrolla herramientas computacionales para las lenguas, clasifica al somalí como "lengua de escasos recursos" y al holandés como "lengua rica o de altos recursos".[133] La importancia de esta diferencia es que las estrategias computacionales aplicadas a uno u otro grupo no son iguales pues las lenguas de escasos recursos presentan más desafíos.

---

[123] Llanes-Ortiz, *Iniciativas digitales para lenguas indígenas*.
[124] Francese, "Language Revitalization through Indigenous Mexican Hip Hop".
[125] Solari Pita et al., "Escucharnos en la pantalla grande: Osankevantite Irira / El Libro de Lila. El proceso de doblaje de películas a lenguas originarias en la comunidad Amazonía - Bajo Urubamba - Cusco – Perú".
[126] Llanes-Ortiz, *Iniciativas digitales para lenguas indígenas*.
[127] Deance Bravo y Troncoso, "TotoOffice".
[128] Llanes-Ortiz, *Iniciativas digitales para lenguas indígenas*.
[129] Dyson, Grant, y Hendriks, *Indigenous People and Mobile Technologies*.
[130] Llanes-Ortiz, *Iniciativas digitales para lenguas indígenas*.

La mayor parte de las lenguas indígenas caen en la categoría de "bajos recursos", pero se refiere únicamente a la escasez de datos textuales u orales en formato digital y no a sus recursos lingüísticos (medios de expresión y comunicación), los cuales son ricos, diversos y altamente efectivos.

Otra característica de las lenguas indígenas es que, a menudo, son ricas en historias orales pero limitadas en registros escritos.[134] La oralidad de estos idiomas enmascara sus sistemas de escritura originales que, aunque en desuso, se ajustan a ellos mejor que la versión latina que requiere de artificios como diferentes tildes, apóstrofes, afijos, infijos y sufijos para representar la fonología y el sentido de una palabra. Esta falta de simplicidad en la escritura ha dificultado su aprendizaje y la producción de tecnología.[135]

---

[131] Joshi et al., "The State and Fate of Linguistic Diversity and Inclusion in the NLP World"; Tonja et al., "NLP Progress in Indigenous Latin American Languages".
[132] Joshi et al., "The State and Fate of Linguistic Diversity and Inclusion in the NLP World".
[133] Mussandi y Wichert, "NLP Tools for African Languages".
[134] Joshi et al., "The State and Fate of Linguistic Diversity and Inclusion in the NLP World".
[135] Joshi et al.; Llanes-Ortiz, *Iniciativas digitales para lenguas indígenas*.

*Desafíos para el PLN*

La diversidad lingüística y los fenómenos tipológicos y fonológicos que caracterizan a las lenguas indígenas hace que la preservación y revitalización de los idiomas nativos no sólo sea de interés para las comunidades indígenas, sino que presente desafíos fascinantes para los investigadores del PLN.[136] Entrenar los sistemas en unas pocas lenguas similares ha llevado a una "cámara de resonancia tipológica", donde muchos fenómenos lingüísticos quedan sin explorar.[137] Considerar la diversidad de escrituras y formas de representación de las lenguas con recursos limitados, facilitará su uso y enriquecerá el campo del PLN al incorporar una gama más amplia de fenómenos lingüísticos y culturales.[138]

Littell y colaboradores[139] resaltan la viabilidad del desarrollo de tecnologías inteligentes para lenguas indígenas, clasificándolas en cuatro categorías que se muestran en la Tabla 5.

---

[136] Littell et al., "Indigenous language technologies in Canada"; Mager et al., "Challenges of language technologies for the indigenous languages of the Americas"; González Santillan, "CS447 Literature Review: Natural Language Processing for Indigenous Languages of the Americas".

[137] Joshi et al., "The State and Fate of Linguistic Diversity and Inclusion in the NLP World"; Marmion, Obata, y Troy, "Community, Identity, Wellbeing".

[138] Joshi et al., "The State and Fate of Linguistic Diversity and Inclusion in the NLP World"; Llanes-Ortiz, *Iniciativas digitales para lenguas indígenas*; Ward, "Qualitative Research in Less Commonly Taught and Endangered Language CALL".

[139] Littell et al., "Indigenous language technologies in Canada".

| Categoría | Descripción | Ejemplos |
|---|---|---|
| Emblemáticas | Requieren grandes volúmenes de datos y son más factibles para lenguas con mejores recursos. | Traducción automática y reconocimiento automático de voz. |
| Prácticas | Ya se han desarrollado y son viables para un mayor número de lenguas indígenas | Diseños de teclado y búsqueda aproximada. |
| En espera | Tienen base tecnológica en ciertos idiomas pero aún no están disponibles para uso amplio. | Revisión ortográfica y generación de paradigmas. |
| Experimentales | Prometedoras pero aún no han demostrado éxito en lenguas indígenas. | Texto predictivo y segmentación de audio. |

Tabla 5. Estado del desarrollo de tecnologías inteligentes para las lenguas indígenas según Littell[140].

Es interesante mencionar que han nacido espacios académicos dedicados a abordar estos desafíos. El NLP Américas (NLPA)[141] y el Africa Language Dataset Challenge (AI4D)[142] son ejemplos notables. En el NLPA, las tareas más investigadas han sido la traducción automática, el análisis morfosintáctico y el reconocimiento de voz,[143] aunque la más reciente edición incluyó por primera vez la creación automática de materiales educativos para el guaraní, el bribri y el maya.[144] Por su parte, el AI4D se enfocó en la creación de datos de calidad en lenguas de Sudáfrica, Ghana y Uganda, que se puedan utilizar en modelos lingüísticos.[145] A la fecha, se han celebrado concursos y hackathones centrados principalmente en las tareas de construcción de corpus, traducción automática y análisis morfosintáctico.[146]

---

[140] Littell et al.
[141] NLPA, "Thrid Workshop on NLP for Indigenous Languages of the Americas (AmericasNLP)".
[142] AI4D, "African Language Dataset Challenge".
[143] Tonja et al., "NLP Progress in Indigenous Latin American Languages".
[144] Chiruzzo et al., "Findings of the AmericasNLP 2024 Shared Task on the Creation of Educational Materials for Indigenous Languages".
[145] Siminyu et al., "AI4D -- African Language Program".
[146] Mussandi y Wichert, "NLP Tools for African Languages".

*Aprendizaje por transferencia: prometedor pero riesgoso*

Los avances en aprendizaje profundo han permitido quitar el foco en la recolección de grandes volúmenes de datos y voltear hacia la optimización de algoritmos.[147] De esta manera se han desarrollado estrategias que están resultando efectivas para las lenguas indígenas, como el aprendizaje por transferencia, con lo cual se han superado limitaciones por falta de datos.[148] Esta técnica posibilita la utilización de modelos pre-entrenados en lenguas de altos recursos para mejorar el rendimiento de tareas en lenguas de escasos recursos. Variantes como el aprendizaje de pocos disparos (pocos datos disponibles) o de disparo cero (ningún ejemplo disponible) son especialmente valiosas en escenarios donde la recolección de datos es difícil o costosa. Ambas se apoyan en información auxiliar, como el contexto semántico, para completar la tarea deseada.[149] Google Translate emplea estas técnicas para incluir lenguas indígenas, muchas de ellas en riesgo, como el n'ko de África, el punjabi de Pakistán, el manés de Irlanda y el maya de México.[150]

Aunque interesante y sorprendente, el aprendizaje por transferencia es un camino riesgoso pues minimiza la participación directa de los hablantes nativos. Esto podría resultar en productos que carecen de la interpretación cultural necesaria. Aún así, es claro que estos enfoques pueden cambiar el destino de las lenguas indígenas,[151] sin embargo, la diversidad lingüística y dialectal sigue planteando desafíos únicos.[152]

---

[147] Littell et al., "Indigenous language technologies in Canada"; Joshi et al., "The State and Fate of Linguistic Diversity and Inclusion in the NLP World"; Mager et al., "Challenges of language technologies for the indigenous languages of the Americas"; González Santillan, "CS447 Literature Review: Natural Language Processing for Indigenous Languages of the Americas".
[148] Littell et al., "Indigenous language technologies in Canada".
[149] Huang et al., "Improving Zero-Shot Cross-Lingual Transfer Learning via Robust Training".
[150] Caswell, "El Traductor de Google incorpora 111 idiomas, siendo su mayor expansión hasta la fecha".
[151] Devlin et al., "BERT"; Joshi et al., "The State and Fate of Linguistic Diversity and Inclusion in the NLP World"; Lample y Conneau, "Cross-lingual Language Model Pretraining"; Pires, Schlinger, y Garrette, "How Multilingual is Multilingual BERT?"
[152] González Santillan, "CS447 Literature Review: Natural Language Processing for Indigenous Languages of the Americas"; Littell et al., "Indigenous language technologies in Canada"; Mager et al., "Challenges of language technologies for the indigenous languages of the Americas".

**CONCLUSIONES Y PROPUESTAS**

En este trabajo hemos dado un panorama general sobre las lenguas indígenas en el mundo, resaltando su diversidad y su estado vital. Hemos identificado a la colonización y a la globalización como las principales causas del desuso y hemos visto que, ante este ataque de unas culturas sobre otras, ha sido necesario legislar los derechos lingüísticos de los pueblos vulnerables. Hicimos un recorrido histórico de las diferentes tecnologías que han sido usadas para estudiar, difundir e intentar rescatar las lenguas y culturas originarias y damos cuenta de que, cuando vienen del exterior, muchas veces tienen el efecto contrario al que buscan; pero, cuando son diseñadas, elaboradas y ejercidas desde adentro de las comunidades, se convierten en instrumentos más genuinos de expresión que van teniendo resonancia a un lado y otro del globo. Al hacer esta revisión han surgido las siguientes reflexiones y propuestas:

- Cada habitante del planeta podría rastrear su ancestría indígena e iniciar un camino de reencuentro con esa identidad olvidada, volviendo a ser conscientes de la coherencia que deberíamos tener con nosotros mismos, con nuestros ancestros, con el territorio que habitamos. Esta sería la campaña de revitalización más grande y, posiblemente, la más efectiva que se haya hecho.
- Los sorprendentes logros actuales de la IA, así como sus peligros potenciales, fueron emergiendo de los datos que la humanidad ha ido depositando en internet usando lenguas y contenidos que promueven el avance científico-tecnológico pero también la dominación. El comportamiento de estos grandes modelos de lenguaje son el mejor ejemplo de que las lenguas son vehículos de epistemologías y valores. Esto nos lleva a preguntarnos ¿qué pasaría si estuvieran incluidas la totalidad de las lenguas humanas con sus sistemas de conocimiento y valores?
- La integración de las lenguas indígenas a la IA eventualmente ocurrirá, pero hay dos

escenarios: una, excluyendo a los hablantes y obteniendo resultados desconectados de la cultura; y la otra, haciendo investigación y diseño tecnológico participativo, con lo que se obtendrían productos útiles y culturalmente pertinentes.

- El enfoque participativo debe ir más allá de considerar a los indígenas como meros proveedores o etiquetadores de datos.En su lugar, debe centrarse en establecer un diálogo de saberes que favorezca las condiciones para que su creatividad y habilidades se expresen, permitiéndoles forjar su propio camino hacia esta nueva era tecnológica.

Cerramos este trabajo enfatizando que, el hecho de que las lenguas indígenas no estén aún integradas en la IA, ofrece la oportunidad de hacerlo con más cuidado. Un verdadero enfoque participativo beneficiaría a las comunidades y enriquecería el panorama tecnológico global. Los estudios sobre diversidad han demostrado que la heterogeneidad en ideas y maneras de pensar lleva a soluciones más eficaces, por lo que la inclusión de valores y sistemas de conocimiento que se alejan del estándar podría resultar en una IA más equitativa y relevante. El día que todas las lenguas existentes estén tecnológicamente representadas, tal vez surja, como una de las grandes habilidades emergentes, la tan buscada alineación de la IA con los mejores valores humanos.

**BIBLIOGRAFÍA**

AI4D. "African Language Dataset Challenge". Artificial Intelligence for Development, el 5 de noviembre de 2019. https://africa.ai4d.ai/blog/african-language-dataset-challenge/.

Ajani, Yusuf Ayodeji, Bolaji David Oladokun, Shuaib Agboola Olarongbe, Margaret Nkechi Amaechi, Nafisa Rabiu, y Musediq Tunji Bashorun. "Revitalizing Indigenous Knowledge Systems via Digital Media Technologies for Sustainability of Indigenous Languages". *Preservation, Digital Technology & Culture* 53, núm. 1 (2024): 35–44.

Alvarez Avila, Abelardo. *Cultura e identidad frente a la globalización*. Barcelona, España: Académica Española, 2016.

Amiel, Sandrine. "¿Quiénes son los pueblos indígenas de Europa y cuáles son sus luchas?" *Euronews*, el 9 de agosto de 2019. https://es.euronews.com/2019/08/09/quienes-son-los-pueblos-indigenas-de-europa-y-cuales-son-sus-luchas.

Angulo, Elena, Christophe Diagne, Liliana Ballesteros-Mejia, Tasnime Adamjy, Danish A. Ahmed, Evgeny Akulov, Achyut K. Banerjee, et al. "Non-English languages enrich scientific knowledge: The example of economic costs of biological invasions". *Science of The Total Environment* 775 (2021): 144441.

Assess Technology. "A qué nos referimos con tecnología". *Assess Technology* (blog). Consultado el 20 de octubre de 2024. https://assess.technology/es/a-que-nos-referimos-con-tecnologia/.

Bahram, Moghaddas. "La enseñanza de lenguas y la cuestión de la cultura en el contexto clásico y digital". *ZENITH Revista Internacional de Investigación Multidisciplinaria* 2, núm. 11 (2013): 272–83.

Banco Mundial. "Latinoamérica Indígena en el Siglo XXI". Washington, D.C., 2015. https://documents1.worldbank.org/curated/en/541651467999959129/pdf/Latinoam%C3%A9rica-ind%C3%ADgena-en-el-siglo-XXI-primera-d%C3%A9cada.pdf.

Batalla, Guillermo Bonfil. "El concepto de indio en América: una categoría de la situación colonial". *Anales de Antropología* 9 (1972).

Bazai, Zia ur Rehman, Syed Abdul Manan, y Stefanie Pillai. "Language Policy and Planning in the Teaching of Native Languages in Pakistan". *Current Issues in Language Planning* 24, núm. 3 (2023): 293–311.

Becerra-Lubies, Rukmini, Simona Mayo, y Aliza Fones. "Revitalización de las lenguas y culturas indígenas: revisión crítica de las políticas de educación preescolar bilingüe en Chile (2007-2016)". *Revista Internacional de Educación Bilingüe y Bilingüismo* 24, núm. 8 (2021): 1147–62.

Bielenberg, Brian. "Indigenous Language Codification: Cultural Effects." En *Annual Stabilizing Indigenous Languages Symposium*, 1–11. ERIC, 1999.

Campbell, Lyle. *American Indian Languages: The Historical Linguistics of Native America*. Revisada. Oxford University Press, 2000.

Cancro, Polly. "The Dark(ish) Side of Digitization: Information Equity and the Digital Divide". *The Serials Librarian* 71, núm. 1 (2016): 57–62.

Castillo Tec, Felipe de Jesús. *U áanalte'il u tsikbalil ts'aak. Manual de frases médicas*. Mérida, México: Instituto para el Desarrollo de la Cultura Maya del Estado de Yucatán, 2013.

Caswell, Isaac. "El Traductor de Google incorpora 111 idiomas, siendo su mayor expansión hasta la fecha". *Google* (blog), el 27 de junio de 2024. https://blog.google/intl/es-419/actualizaciones-de-producto/informacion/el-traductor-de-google-incorpora-111-idiomas-siendo-su-mayor-expansion-hasta-la-fecha/.

Cazden, Courtney B. "Sustaining Indigenous Languages in Cyberspace". *Nurturing Native Languages. Northern Arizona University*, 2003, 1–7.

Chávez Ángeles, Manuel Gerardo, y Joselito Fernández Tapia. "Etnografía cuantitativa. Revitalización lingüística y difusión de las tecnologías digitales en municipios de Oaxaca, México". *Alteridades* 30, núm. 59 (2020): 111–21.

Chemla, Karine. *History of Science, History of Text*. Springer Science & Business Media, 2005.

Chiocca, Emmanuelle S. "Language Endangerment: Diversity and Specificities of Native American Languages of Oklahoma". En *Handbook of the Changing World Language Map*, 1–20. Cham: Springer International Publishing, 2018.

Chiruzzo, Luis, Pavel Denisov, Alejandro Molina-Villegas, Silvia Fernandez-Sabido, Rolando Coto-Solano, Marvin Agüero-Torales, Aldo Alvarez, et al. "Findings of the AmericasNLP 2024 Shared Task on the Creation of Educational Materials for Indigenous Languages". En *Proceedings of the 4th Workshop on Natural Language Processing for Indigenous Languages of the Americas (AmericasNLP 2024)*, 224–35. Mexico City, Mexico: Association for Computational Linguistics, 2024.

Choudhury, Rumman. *De todas formas, tu opinión no importa. La violencia de género facilitada por la tecnología en la era de la IA generativa*. París, Francia: UNESCO, 2024.

Corbetta, Silvina, Carlos Bonetti, Fernando Bustamante, y Albano Vergara Parra. "Educación intercultural bilingüe y enfoque de interculturalidad en los sistemas educativos latinoamericanos. Avances y desafíos". Comisión Económica para América Latina y el Caribe (CEPAL), 2018.

Corona, Ignacio. "La identidad indígena como identidad urbana. Un abordaje descolonial a las crónicas de Ana Matías Rendón". *Latinoamérica. Revista de Estudios Latinoamericanos* 1, núm. 76 (2023): 23–51.

Council of Europe. "European Day of Languages". Consultado el 26 de febrero de 2025. https://edl.ecml.at/Facts/LanguageFacts/tabid/1859/language/en-GB/Default.aspx.

Day, David. *Conquista. Una nueva historia del mundo moderno*. Barcelona, España: Crítica, 2006.

Deance Bravo y Troncoso, Iván. "TotoOffice: Experiencias interculturales en torno a la lengua y la tecnología." En *Los territorios discursivos en América Latina-Interculturalidad,Comunicación e Identidad*, Primera., 98–115. Quito, Ecuador: Ediciones Ciespal, 2017.

Department of Maori Affairs, New Zeland. Maori Language Act, Pub. L. No. 176 (1987).

Devlin, Jacob, Ming-Wei Chang, Kenton Lee, y Kristina Toutanova. "BERT: Pre-training of Deep Bidirectional Transformers for Language Understanding". En *Proceedings of the 2019 Conference of the North American Chapter of the Association for Computational Linguistics*, 1:4171–86. Human Language Technologies. ACL, 2019.

DOF, México. Ley General de Derechos Lingüísticos de los Pueblos Indígenas, Diario Oficial de la Federación § (2003).

———. Programa Especial de los Pueblos Indígenas, Diario Oficial de la Federación § (2014).

DOGC, España. Ley de política lingüística de Catalunya (1998).

Dooly, Melinda, y Anna Comas-Quinn. "Access to technology and social justice". En *La enseñanza del español mediada por tecnología*, 1a ed., 24–47. London: Routledge, 2023.

Dyson, Laurel Evelyn, Stephen Grant, y Max Hendriks. *Indigenous People and Mobile Technologies*. 1st ed. Vol. 1. Nueva York: Routledge, Taylor & Francis Group, 2016.

Dyson, Laurel Evelyn, Max Hendriks, y Stephen Grant. *Information Technology and Indigenous People*. IGI Global, 2007.

Fernández-Sabido, Silvia, Yoly Palomo-Carrillo, Rafael Burgos-Villanueva, y Romeo de Coss. "Comparative Study of Two Blue Pigments from the Maya Region of Yucatan". *MRS Online Proceedings Library* 1374, núm. 1 (2012): 115–23.

Francese, Jonah. "Language Revitalization through Indigenous Mexican Hip Hop: Building towards an Indigenous Hip Hop Futurism". *Journal of Multilingual and Multicultural Development* 45, núm. 1 (2023): 9–21.

Francour, Daisee. "Revitalizing the Oneida Language through Indigenous Language Immersion". *Cultural Survival*, el 31 de agosto de 2022. https://www.culturalsurvival.org/publications/cultural-survival-quarterly/revitalizing-oneida-language-through-indigenous-language.

Garcés Velásquez, Luis Fernando. "Las comunidades virtuales del quichua ecuatoriano: revalorizando la lengua en un espacio apropiado". *Tellus* 20, núm. 43 (2021): 59–75.

Gobierno de Bolivia. Constitución Política del Estado de Bolivia (2009).

Gobierno de Ecuador. Constitución de Ecuador (2008).

Gobierno de México. "Las radios comunitarias y su labor en el desarrollo rural", 2023. http://www.gob.mx/siap/es/articulos/las-radios-comunitarias-y-su-labor-en-el-desarrollo-rural.

González Santillan, Diana. "CS447 Literature Review: Natural Language Processing for Indigenous Languages of the Americas", 2021.

Gutiérrez Fonseca, Itza Nahomy. "Las epidemias del México prehispánico: un breve recorrido histórico". *Revista de Medicina y Cine* 16, núm. e (2021): 237–45.

Hallett, Darcy, Michael J. Chandler, y Christopher E. Lalonde. "Aboriginal language knowledge and youth suicide". *Cognitive Development* 22, núm. 3 (2007): 392–99.

Hamel, Rainer Enrique. "Derechos lingüísticos como derechos humanos: debates y perspectivas". *Alteridades* 5, núm. 10 (1995): 11–23.

Holmes, Frederic, Jürgen Renn, y Hans-Jörg Rheinberger. *Reworking the Bench: Research Notebooks in the History of Science*, 2003.

Huang, Kuan-Hao, Wasi Uddin Ahmad, Nanyun Peng, y Kai-Wei Chang. "Improving Zero-Shot

Cross-Lingual Transfer Learning via Robust Training". arXiv, 2021.

Ibáñez Blancas, Nicolás, Edgar Isch L., Daniel Panario, Ofelia Gutierrez, y Ángela Zambrano C. "El cambio climático y los conocimientos tradicionales, miradas desde Sudamérica". *Terra Nueva Etapa* XXXVI, núm. 59 (2020).

IIWGIA, The International Work Group for Indigenous Affairs. "IWGIA Annual Report", 2023.

INEGI. "Encuesta Nacional sobre Disponibilidad y Uso de Tecnologías de la Información en los Hogares 2023. ENDUTIH." México, 2024.

ITACAT. "Las lenguas de Asia". Agencia de comunicación intercultural de Cataluña, 2011. https://www.itacat.info/2011/09/las-lenguas-de-asia.html.

ITF. "Estudio de los determinantes de la conectividad y despliegue de redes móviles en México". Estadístico. México: Instituto Federal de Telecomunicaciones, 2024.

Iturriaga Acevedo, Eugenia, y Jaime López Reyes. *Pueblos indígenas frente al racismo mexicano*. 1era ed. Ciudad de México, México: Contramarea Editorial - UNAM, 2020.

Ivers, Laura. "Pueblos indígenas: Panorama general". *World Bank*, 2023.

Janetsky, Megan. "Estas parteras tradicionales combinan la herencia maya con la medicina occidental para salvar vidas". National Geographic, el 8 de abril de 2022. https://www.nationalgeographicla.com/historia/2022/04/estas-parteras-tradicionales-combinan-la-herencia-maya-con-la-medicina-occidental-para-salvar-vidas.

Johansson, Patrick. "Miguel León-Portilla y el mundo indígena". En *Su aliento, palabra: homenaje a Miguel León-Portilla*, 205–20. México: UniverUNAM, Colegio Nacional, INAH, 1997.

José-Yacamán, M., Luis Rendón, J. Arenas, y Mari Carmen Serra Puche. "Maya Blue Paint: An Ancient Nanostructured Material". *Science* 273, núm. 5272 (1996): 223–25.

Joshi, Pratik, Sebastin Santy, Amar Budhiraja, Kalika Bali, y Monojit Choudhury. "The State and Fate of Linguistic Diversity and Inclusion in the NLP World". En *Proceedings of the 58th Annual Meeting of the Association for Computational Linguistics*, 6282–93. ACL, 2020.

Klein, Ursula. "Paper tools in experimental cultures". *Studies in History and Philosophy of Science Part A* 32, núm. 2 (2001): 265–302.

Koller, Eve, y Malayah Thompson. "The Representation of Indigenous Languages of Oceania in Academic Publications". *Publications* 9, núm. 2 (2021): 1–13.

Lample, Guillaume, y Alexis Conneau. "Cross-lingual Language Model Pretraining". arXiv, 2019.

Langlois, Jills. "Las muertes de ancianos por la COVID-19 ponen en peligro los idiomas indígenas". National Geographic, el 16 de noviembre de 2020. https://www.nationalgeographic.es/historia/2020/11/muertes-de-ancianos-por-covid-19-ponen-en-peligro-idiomas-indigenas.

Lenoir, Timothy. *Inscribing Science*. Stanford University Press, 1998.

Littell, Patrick, Anna Kazantseva, Roland Kuhn, Aidan Pine, Antti Arppe, Christopher Cox, y Marie-Odile Junker. "Indigenous language technologies in Canada: Assessment, challenges, and successes". En *Proceedings of the 27th International Conference on Computational Linguistics*, 2620–32, 2018.

Llanes-Ortiz, Genner. *Iniciativas digitales para lenguas indígenas*. París, Francia: UNESCO/Global Voices, 2023.

Lopez, Luis, y Carlos Callapa. "Situación general de las lenguas indígenas y políticas gubernamentales en América Latina y el Caribe", 2019.

Mager, Manuel, Rajat Bhatnagar, Graham Neubig, Ngoc Thang Vu, y Katharina Kann. "Neural Machine Translation for the Indigenous Languages of the Americas: An Introduction".

arXiv, 2023.

Mager, Manuel, Ximena Gutierrez-Vasques, Gerardo Sierra, y Ivan Meza-Ruiz. "Challenges of language technologies for the indigenous languages of the Americas". En *Proceedings of the 27th International Conference on Computational Linguistics*, 55–69. Santa Fe, New Mexico, USA: ACL, 2018.

Mager, Manuel, y Iván Meza. "Retos en construcción de traductores automáticos para lenguas indígenas de México". *Digital Scholarship in the Humanities* 36, núm. Supplement_1 (2021): i43–48.

Makgopa, Mokgale. "Implications of the Fourth Industrial Revolution (4IR) on the Development of Indigenous Languages of South Africa: Challenges and Opportunities". En *Handbook of Research on Protecting and Managing Global Indigenous Knowledge Systems*, 152–61. IGI Global, 2022.

Maldonado Rivera, M. R., F. Blanco Sánchez, y G. Ramiro Reyes. "Las tecnologías de la información y la comunicación (TIC) fortalecen la preservación de las lenguas indígenas". *TECTZAPIC Revista de divulgación científica y tecnológica* 4, núm. 1 (2018): 49–55.

Mallén Rivera, Carlos. "La ciencia en el México colonial e independiente". *Revista mexicana de ciencias forestales* 3, núm. 9 (2012): 03–09.

Marino-Jiménez, Mauro, Fany Olinda Rojas-Noa, y Karina Natalia Sullón-Acosta,. "Lenguas indígenas: un sistema de educación y preservación a través de la tecnología, las presiones institucionales y el pensamiento sistémico". En *Tecnologías educativas y estrategias didácticas*, 1a ed., 585–96. Málaga, España: UMA Editorial, 2020.

Marmion, Doug, Kazuko Obata, y Jakelin Troy. "Community, Identity, Wellbeing: The Report of the Second National Indigenous Languages Survey". National Indigenous Languages Survey, 2014.

Martín Briceño, Patrizia, y Fidencio Briceño Chel. *Manual de Comunicación para Médicos. Español-maya.* Mérida México: Universidad Autónoma de Yucatán, 2013.

Marwala, Tshilidzi. *Closing the Gap: The Fourth Industrial Revolution in Africa*. Johannesburg, South Africa, 2020.

———. "The Fourth Industrial Revolution has arrived. Comments on Moll (S Afr J Sei. 2023;119(1/2), Art. #12916)". *South African Journal of Science* 119, núm. 1/2 (2023).

Mateos Cortés, Laura Selene, Gunther Dietz, y Gunther Dietz. "Universidades interculturales en México: balance crítico de la primera década". *Revista mexicana de investigación educativa* 21, núm. 70 (2016): 683–90.

Mcquown, Norman A. "The indigenous languages of latin america". *American Anthropologist*, 1955.

Meyer, Julien, Laure Dentel, y Frank Seifart. "A methodology for the study of rhythm in drummed forms of languages: application to Bora Manguaré of Amazon." En *INTERSPEECH*, 687–90, 2012.

Mkontwana, Phinda, y Doctor Sundani Ndivhuwo. "The Role of South African Community Radio Stations in Promoting South African Indigenous Languages: A Systematic Review". *International Journal of Social Science Research and Review* 6, núm. 10 (2010): 179–89.

MMMMarino-Jiménez, Mauro, Fany Olinda Rojas-Noa, y Karina Natalia Sullón-Acosta,. "Lenguas indígenas: un sistema de educación y preservación a través de la tecnología, las presiones institucionales y el pensamiento sistémico". En *Tecnologías educativas y estrategias didácticas*, 1a ed., 585–96. Málaga, España: UMA Editorial, 2020.

https://monografias.uma.es/index.php/mumaed/catalog/book/71.

Montejo Cruz, Oscar, José Bastiani Gómez, y Segundo Jordán Orantes Orantes. "Experiencia digital en la enseñanza del Ch'ol en Chiapas, México". *Revista Senderos Pedagógicos* 10, núm. 1 (2019): 145–62.

Moodley, Maglin, y Reuben Dlamini. "Experiences and Attitudes of Setswana Speaking Teachers in Using an Indigenous African Language on an Online Assessment Platform". *South African Journal of Education* 41, núm. Supplement 1 (2021): S1–11.

Muñoz-Basols, Javier, Mara Fuertes Gutiérrez, y Luis Cerezo. *La enseñanza del español mediada por tecnología; de la justicia social a la Inteligencia Artificial (IA)*. Taylor & Francis, 2024.

Mussandi, Joaquim, y Andreas Wichert. "NLP Tools for African Languages: Overview". En *Proceedings of the 16th International Conference on Computational Processing of Portuguese*, 73–82. Santiago de Compostela, Galicia/Spain: Association for Computational Lingustics, 2024.

NLPA. "Thrid Workshop on NLP for Indigenous Languages of the Americas (AmericasNLP)", 2021. http://turing.iimas.unam.mx/americasnlp/index.html.

OIT. "Convenio C169 - Convenio sobre pueblos indígenas y tribales", 1989. https://normlex.ilo.org/dyn/nrmlx_es/f?p=NORMLEXPUB:12100:0::NO::P12100_ILO_CODE:C169.

Omojola, Oladokun. "English-Oriented ICTs and Ethnic Language Survival Strategies in Africa". *Global Media Journal: African Edition* 3, núm. 1 (2009).

Omusonga, Tony Okwach, Pius Simiyu, y Daniel K. Chesaro. "Learning Indigenous Languages in Public Primary Schools in Kenya: Overview, Problems and the Way Forward". *European Journal of Education Studies* 10, núm. 12 (2023).

ONU. "Pueblos indígenas | Naciones Unidas". United Nations, s.f. https://www.un.org/es/fight-racism/vulnerable-groups/indigenous-peoples.

ONU, Departamento de Información Pública. "Foro permanente para las cuestiones indígenas - Documento de antecedentes", 2019. https://www.un.org/es/events/indigenousday/assets/pdf/Backgrounder-Languages-Spanish%202019.pdf.

Peterson, Tyler, y Ofelia Zepeda. "Towards an Indigenously-informed Model for Assessing the Vitality of Native American Languages: a Southern Arizona Pilot Project". *Language Documentation & Conservation Special Publication* 27 (2022): 136–54.

Pillajo Zambrano, Leandro Damián. "Producción del Encuentro Nacional de Activistas Digitales de Lengua Indígena (ENADLI)". PhD Thesis, Universitat Politècnica de València, 2019.

Pires, Telmo, Eva Schlinger, y Dan Garrette. "How Multilingual is Multilingual BERT?" En *Proceedings of the 57th Annual Meeting of the Association for Computational Linguistics*, 4996–5001. Florence, Italy: Association for Computational Linguistics, 2019.

Reyhner, Jon, Louise Lockard, y Joseph Martin. "Revitalizing Indigenous Languages Challenges and Opportunities". Northern Arizona University, 2024.

Ríos Hernández, Iván. "El lenguaje: herramienta de reconstrucción del pensamiento". *Razón y Palabra*, núm. 72 (2010).

Roche, Gerald, Madoka Hammine, Jesus Federico C. Hernandez, y Jess Kruk. "The Politics of Fear and the Suppression of Indigenous Language Activism in Asia: Prospects for the United Nations' Decade of Indigenous Languages". *State Crime Journal* 12, núm. 1 (2023). https://scienceopen.com/hosted-document?doi=10.13169/statecrime.12.1.0029.

Rodríguez Caguana, Adriana. *El largo camino del Taki Unkuy: los derechos lingüísticos y culturales de los pueblos indígenas del Ecuador*. Huaponi Ediciones, 2017.

Ronquillo, Cynthia Gabriela. "El derecho de los adultos mayores a acceder al consumo mediado por la tecnología digital en Argentina como eje de la concreción de sus derechos humanos". Tesis doctoral, Universidad Nacional de La Matanza, 2023.

Rosa, Jonathan, y Nelson Flores. "Decolonization, Language, and Race in Applied Linguistics and Social Justice". *Applied Linguistics* 42, núm. 6 (2021): 1162–67.

Sandoval-Forero, Eduardo Andrés. "Hitos Demográficos del Siglo XXI: Población Indígena". *Espacio Abierto* 27, núm. 3 (2018): 203–7.

Secretaría de Cultura. "Tipografía para lenguas indígenas, herramienta de conservación", 2016. https://www.gob.mx/cultura/prensa/tipografia-para-lenguas-indigenas-herramienta-de-conservacion?state=published.

SEP, Secretaría de Educación Pública. "Coordinación General de Educación Intercultural y Bilingüe (CGEIB)", 2015. https://dgeiib.basica.sep.gob.mx/files/fondo-editorial/educacion-intercultural/cgeib_00001.pdf.

Siminyu, Kathleen, Godson Kalipe, Davor Orlic, Jade Abbott, Vukosi Marivate, Sackey Freshia, Prateek Sibal, et al. "AI4D -- African Language Program". arXiv, 2021.

Solari Pita, Helder, Ángela de la Torre Tupayachi, Marcela Rincón González, y Alexander Muñoz Ramírez. "Escucharnos en la pantalla grande: Osankevantite Irira / El Libro de Lila. El proceso de doblaje de películas a lenguas originarias en la comunidad Amazonía - Bajo Urubamba - Cusco – Perú". En *Experiencias en revitalización y fortalecimiento de lenguas indígenas en Latinoamérica : Argentina - Bolivia - Chile - Ecuador - Guatemala - México - Perú*, 1a ed., 296. Cochabamba, Bolivia: Funproeib Andes, 2021.

South African Government. The Constitution of the Republic of South Africa (1997). https://www.gov.za/documents/constitution/constitution-republic-south-africa-04-feb-1997.

Stalder, Félix. *Manuel Castells: The Theory of the Network Society*. 1a ed. Key Contemporary Thinkers | Key Contemporary Thinkers. Cambridge: Polity Press, 2006.

Teubner, Timm, Christoph M. Flath, Christof Weinhardt, Wil van der Aalst, y Oliver Hinz. "Welcome to the Era of ChatGPT et Al." *Business & Information Systems Engineering* 65, núm. 2 (2023): 95–101.

Tierney, Robert J., Kadriye Ercikan, y Fazal Rizvi, eds. "Diversity, Democracy, and Social Justice in Education: Africa". En *International Encyclopedia of Education*, 4a ed., 2006. https://www.sciencedirect.com/topics/social-sciences/african-languages.

Tonja, Atnafu, Fazlourrahman Balouchzahi, Sabur Butt, Olga Kolesnikova, Hector Ceballos, Alexander Gelbukh, y Thamar Solorio. "NLP Progress in Indigenous Latin American Languages". En *Proceedings of the 2024 Conference of the North American Chapter of the Association for Computational Linguistics: Human Language Technologies*, 6972–87. Mexico City, Mexico: ACL, 2024.

Tshifhumulo, Rendani, y Tshimangadzo Justice Makhanikhe. *Handbook of Research on Protecting and Managing Global Indigenous Knowledge Systems*. IGI Global, 2021.

Uekusa, Shinya. "Disaster Linguicism: Linguistic Minorities in Disasters". *Language in Society* 48, núm. 3 (2019): 353–75.

UNESCO. "The World Atlas of Languages", 2024. https://en.wal.unesco.org/world-atlas-languages.

Valladolid, Julio R., y P.B. Sandoval Chayña. "Señas ancestrales como indicadores biológicos de alerta temprana". Programa Mundial de Alimentos, Naciones Unidas, 2007. https://es.wfp.org/publicaciones/peru-senas-ancestrales-como-indicadores-biologicos-de-alerta-temprana.

Vallejo, Claudia, y Melinda Dooly. "Plurilingualism and translanguaging: emergent approaches and shared concerns. Introduction to the special issue". *International Journal of Bilingual Education and Bilingualism* 23, núm. 1 (2020): 1–16.

Ward, Monica. "Qualitative Research in Less Commonly Taught and Endangered Language CALL". *Language Learning & Technology*, 2018.

*What Are Minority, Indigenous, and Endangered Languages?* Lindsay Does Languages, 2019. https://www.youtube.com/watch?v=xxAeuoSgN5c.

WOLACO, Asociación. "Familias de lenguas en Oceanía". Programa Sorosoro. Consultado el 14 de octubre de 2024. https://www.sorosoro.org/es/familias-de-lenguas-en-oceania/.

Yakovlevna Unarova, Vilena. "Design Of A Developing Speech Environment Using Native Languages Of Indigenous Minorities", 1654–63, 2021.